\documentclass[superscriptaddress,showpacs,preprintnumbers,amsmath,amssymb,prc,twocolumn, nofootinbib]{revtex4-2}
\usepackage{color}
\usepackage{graphicx}
\usepackage{caption}
\usepackage{subcaption}
\usepackage{tabularx} 
\usepackage{amsmath}  
\usepackage{scrextend}
\usepackage{hyperref}
\usepackage{dcolumn}
\usepackage{bm}
\usepackage{IEEEtrantools}

\DeclareMathOperator\erfc{erfc}

\newcommand{\overbar}[1]{\mkern 1.5mu\overline{\mkern-1.5mu#1\mkern-1.5mu}\mkern 1.5mu}

\newcommand{\RomanNumeralCaps}[1]
{\MakeUppercase{\romannumeral #1}}

\begin{document}

\preprint{LA-UR-20-23382}

\title{Effect of an electric field on liquid helium scintillation produced by fast
  electrons}

\author{N.~S.~Phan}
\email{nphan@lanl.gov}
\affiliation{Los Alamos National Laboratory, Los Alamos, New Mexico 87545, USA}

\author{V.~Cianciolo}
\affiliation{Oak Ridge National Laboratory, Oak Ridge, Tennessee 37831 USA}

\author{S.~M.~Clayton}
\affiliation{Los Alamos National Laboratory, Los Alamos, New Mexico 87545, USA}

\author{S.~A.~Currie}
\affiliation{Los Alamos National Laboratory, Los Alamos, New Mexico 87545, USA}

\author{R.~Dipert}
\affiliation{Department of Physics, Arizona State University, Tempe, Arizona 85287, USA}

\author{T.~M.~Ito}
\email{ito@lanl.gov}
\affiliation{Los Alamos National Laboratory, Los Alamos, New Mexico 87545, USA}

\author{S.~W.~T.~MacDonald}
\affiliation{Los Alamos National Laboratory, Los Alamos, New Mexico 87545, USA}

\author{C.~M.~O'Shaughnessy}
\affiliation{Los Alamos National Laboratory, Los Alamos, New Mexico 87545, USA}

\author{J.~C.~Ramsey}
\affiliation{Oak Ridge National Laboratory, Oak Ridge, Tennessee 37831 USA}

\author{G.~M.~Seidel}
\email{george\_seidel@brown.edu}
\affiliation{Department of Physics, Brown University, Providence, Rhode Island 02912, USA}

\author{E.~Smith}
\affiliation{Los Alamos National Laboratory, Los Alamos, New Mexico 87545, USA}

\author{E.~Tang}
\affiliation{Los Alamos National Laboratory, Los Alamos, New Mexico 87545, USA}

\author{Z.~Tang}
\affiliation{Los Alamos National Laboratory, Los Alamos, New Mexico 87545, USA}

\author{W.~Yao}
\affiliation{Oak Ridge National Laboratory, Oak Ridge, Tennessee 37831 USA}

\date{\today}

\begin{abstract}
	
The dependence on applied electric field ($0 - 40$~kV/cm) of the scintillation light produced by fast electrons and $\alpha$ particles stopped in liquid helium in the temperature range of 0.44 K to 3.12 K is reported.  For both types of particles, the reduction in the intensity of the scintillation signal due to the applied field exhibits an apparent temperature dependence.  Using an approximate solution of the Debye-Smoluchowski equation, we show that the apparent temperature dependence for electrons can be explained by the time required for geminate pairs to recombine relative to the detector signal integration time.  This finding indicates that the spatial distribution of secondary electrons with respect to their geminate partners possesses a heavy, non-Gaussian tail at larger separations, and has a dependence on the energy of the primary ionization electron.  We discuss the potential application of this result to pulse shape analysis for particle detection and discrimination.

\end{abstract}

\pacs{Valid PACS appear here}
                             
\maketitle

\section{\label{sec:intro}Introduction}
Scintillation of liquid helium\footnote{Here the term ``liquid helium"
refers to either liquid $^4$He or liquid helium with the natural
isotropic abundance.} (LHe) in response to the passage of
charged particles was discovered in the late
1950s~\cite{THO59,FLE59}. Since then extensive studies have been carried out to
illuminate the behavior of ions and neutrals in this unique substance~\cite{MayerReif1958, Mos63, Hereford1966, Mel69, Man71, Rob71, Rob72, Rob73, Kane1963, Miller1964, Dunscombe1971}.  More recently, there has been renewed interest in studying liquid
helium scintillation because of the potential application of LHe as a
particle detector and/or a target material in which to conduct
nuclear, particle, and astroparticle physics experiments~\cite{ Adams1998, AdamsThesis, HabichtThesis, McKinsey1999, McKinsey2003, McKinsey2004, GUO12, Ito2012, Seidel2014}. These
experiments include solar neutrino detection~\cite{LAN87,LAN88,HUA08},
a search for the permanent electric dipole moment of the
neutron~\cite{GOL94,ITO07,SNSCollab2019}, measurement of the free neutron
lifetime~\cite{HUF00}, and detection of light dark-matter
particles~\cite{GUO13,ITO13,MAR17,Hertel2019}.

These wide-ranging applications are motivated by one or more of the following
unique properties of LHe:
(1) LHe can be made with very high purity~\cite{MUMM2016}. Apart from
$^3$He, the only solute of any significance in liquid
$^4$He is hydrogen, which has a solubility of $10^{-14}$ at
1~K~\cite{JEW79}.
(2) Superfluid helium provides multiple signal channels, including
electric charge, prompt scintillation, delayed scintillation, and
elementary excitations, allowing for particle identification.
(3) The low mass of $^4$He provides relatively good kinematic matching
to search for GeV-scale dark-matter particles. 
(4) Superfluid helium can be used to produce ultracold neutrons via
downscattering~\cite{GOL77}.
(5) LHe is a good electrical insulator~\cite{ITO16}.
(6) Dissolved spin-polarized $^3$He atoms can serve as a cohabiting
magnetometer~\cite{GOL94}.
(7) Dissolved $^3$He atoms allow neutron detection via
the reaction $^3$He($n$,$p$)$^3$H, whose reaction products produce
scintillation light in LHe~\cite{GOL94}. 

The passage of a charged particle in LHe deposits energy
into the medium by ionizing and exciting helium atoms. Ionization
creates electrons and ions, which then thermalize with the LHe. The electron subsequently forms a ``bubble'' in the liquid,
pushing away surrounding helium atoms as a consequence of Pauli
exclusion. The He$^+$ ion, on the other hand, forms a ``snowball''
by attracting surrounding helium atoms. The bubbles and snowballs
recombine to form excited helium molecules (excimers). The excited
atoms also form excimers by attracting nearby helium atoms. These
excimers are formed in singlet and triplet states. The lowest
singlet-state molecule radiatively decays in less than 10~ns to
the (unbound) ground state, emitting an $\approx 16$-eV (80~nm)
extreme-ultraviolet (EUV) photon and generating the prompt component
of the LHe scintillation. The triplet state molecule, on the other
hand, has a lifetime of $\approx 13$~s in LHe~\cite{McKinsey1999}. In a
high-excitation-density environment, the triplet-state excimers can
undergo the Penning ionization process,
\begin{equation}
\label{eq:penning1}
{\rm He}_2^* + {\rm He}_2^* \rightarrow 3{\rm He} + {\rm He}^+ + e^-,
\end{equation}
or
\begin{equation}
\label{eq:penning2}
{\rm He}_2^* + {\rm He}_2^* \rightarrow 2{\rm He} + {\rm He}_2^{\;+} + e^-.
\end{equation}
If a singlet excimer is formed as a result of Penning ionization, then it
produces the delayed scintillation component (sometimes referred to as
``afterpulses'').

The ionization density, and as a consequence, the charge distribution
about the particle track, depends on the type of charged particle. For
example, a 5.5~MeV $\alpha$ particle, such as those from an $^{241}$Am
source, has a range of $\approx 0.3 $~mm in superfluid helium ($\rho =
0.145$~g/cm$^3$)~\cite{ASTAR}. As a result, it produces dense,
interpenetrating columns of positive and negative charges. The radius of the columns was
estimated to be $\approx 60$~nm~\cite{Ito2012}. The proton and triton from
the $^3$He($n$,$p$)$^3$H reaction, which share a kinetic energy of
760~keV, produce similar columns of charges, but with a lower
density~\cite{Ito2012}.

On the other hand, a 364-keV electron, such as those from $^{113}$Sn,
has a range of $\approx 7$~mm in superfluid helium~\cite{ESTAR}. With a
$W$ value, defined as the average energy loss by the incident particle per ion pair formed, of 43~eV~\cite{JES55}, the average separation of ionization events is $\approx 840$~nm, whereas the average separation between the
electron bubble and the helium snowball from an electron-ion pair
after they thermalized is $\approx 40$~nm~\cite{Seidel2014}. As a result, the thermalized ions from electron tracks are
most likely to recombine with their partners, a situation referred to as geminate recombination.

In both cases, if an electric field is applied, then the recombination
process is suppressed since some fraction of the charges that would have
otherwise recombined are pulled apart, resulting in a reduced scintillation yield
for both the prompt and delayed components. The separated charges are
collected at the electrodes used to apply the electric field. Importantly, 
only the component of scintillation light that results from
recombination is affected by the electric field. The component that
results from excited atoms is left unaffected.

To describe recombination under an applied electric field, Jaffe's
columnar theory of recombination~\cite{JAF13,KRA52} is most applicable to
the high ionization density case, such as $\alpha$ particles and the
protons and tritons from the $^3$He($n$,$p$)$^3$H reaction. On the
other hand, Onsager developed a theory to describe geminate
recombination~\cite{Onsager1938}, taking into consideration the applied external field
and the Coulomb attraction between the two charges, as well as 
thermal diffusion. The length scale of most relevance in this analysis 
is the so-called Onsager radius, $R=e^2/(4\pi \epsilon k_B T)$, 
which is the separation distance between the two charges where the Coulomb 
potential is comparable to the thermal energy.  In LHe, this radius
is larger than 3.7~$\mu$m, and therefore thermal diffusion can be ignored 
for moderate or higher fields~\cite{Seidel2014}.

In applications of LHe to particle detection, 
an electric field is often applied either to collect charge as
one of the signal channels or to satisfy other experimental
requirements. Therefore, it is of general interest to understand the
light/charge response of LHe as a function of the strength
of the applied electric field.  We have measured the effect of an electric
field on LHe scintillation produced by fast, monoenergetic electrons with a kinetic energy of 364~keV from a $^{113}$Sn source for electric fields up to 40~kV/cm and helium temperatures
between 0.44~K and 3.12~K at a pressure of 600~Torr. This work was
conducted as part of the effort to develop an experiment to search for
the permanent electric dipole moment of the
neutron~\cite{GOL94, ITO07, SNSCollab2019}.  Previous works related to this effort as well as that for the measurement of the neutron lifetime are found in Ref.~\cite{HabichtThesis, McKinsey1999, McKinsey2003, McKinsey2004}.  Our group has previously measured the
effect of an electric field on LHe scintillation produced by
$\alpha$ particles~\cite{Ito2012}. We have also reported on the field
dependence of the ionization current from electrons in LHe
and the resulting determination of the charge thermalization distribution
~\cite{Seidel2014}. Guo {\it et al.}~\cite{GUO12} have reported on a
measurement of the effect of an electric field on LHe
scintillation produced by electrons up to an electric field of 5~kV/cm at
a single temperature of 1.5~K. The work presented in this paper
significantly expands on both the electric field and temperature
ranges. As described below, our data not only
provide important information for using LHe for particle detection
in a wide range of physics experiments but also give new insights into
the process of geminate charge recombination and its time and temperature dependence.  Furthermore, it may be possible to apply these findings to particle detection and identification through pulse shape analysis.

This paper is organized as follows. Section~\ref{sec:exp}
describes the experimental apparatus and
methods. Section~\ref{sec:data} presents the data and the details of
the analysis method employed. Section~\ref{sec:discussion} discusses
the results and their interpretation.

\section{\label{sec:exp}Experimental apparatus and procedure}

\begin{figure*}
	\includegraphics[width=\textwidth]{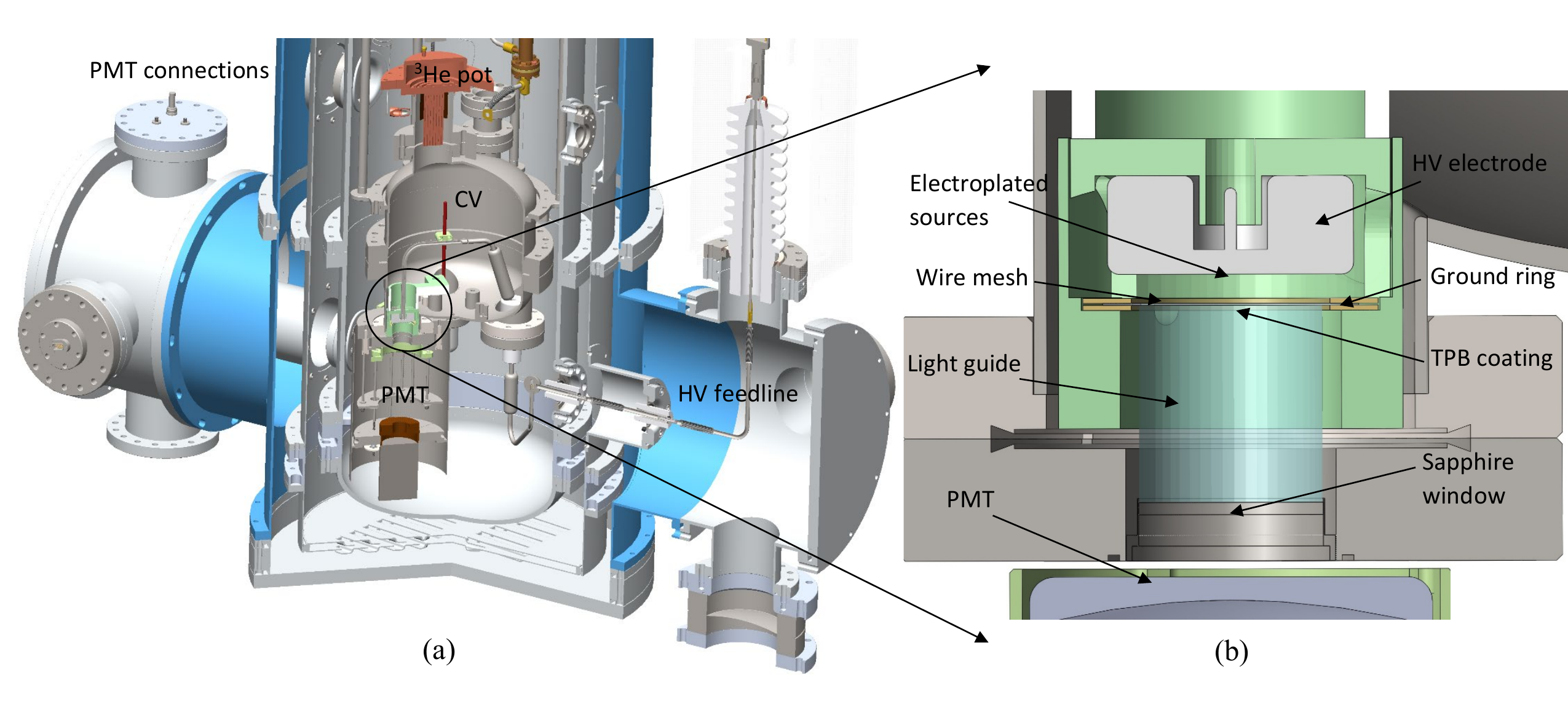}
	\caption{(a) A diagram of the MSHV Test Apparatus and (b) an enlarged view of the light detection system.}
	\label{fig:setup}
\end{figure*}

\subsection{\label{sec:apparatus} Apparatus}
The Medium-Scale High Voltage (MSHV) Test Apparatus~\cite{ITO16} was
used to perform the experiment presented in this paper. The MSHV Test
Apparatus is a cryogenic apparatus designed and constructed to study
electrical breakdown in LHe at temperatures as low as 0.4~K
for pressures between the saturated vapor pressure and 600~Torr. In
this apparatus, the 6-l Central Volume (CV), which can accommodate
a pair of electrodes as large as 12~cm in diameter, is cooled down to
0.4~K using a $^3$He refrigerator. A potential of up to $\pm50$~kV can
be applied to each of the electrodes. For this experiment, we replaced
the MSHV electrode system with the assembly depicted in
Fig.~\ref{fig:setup}. This assembly consists of (1) a high-voltage
electrode 3.18~cm in diameter, on which $^{113}$Sn and
$^{241}$Am radioactive sources were electroplated; (2) an electropolished wire mesh that serves as a
ground electrode; (3) a cylindrical light guide made of poly(methyl methacrylate)
(PMMA), 2.54~cm in diameter and 2.54~cm in
length, whose end facing the wire mesh was coated with vacuum
evaporated tetraphenyl butadiene (TPB), which converts the 80-nm EUV
light to 400-nm visible light; and (4) a G10 structure to hold these
components together. As shown in Fig.~\ref{fig:setup}, this assembly was mounted
on one of the ports of the MSHV CV. The 400-nm light from the TPB-coated 
end of the light guide is guided to the other end, through a
sapphire viewport, and, finally, to a Hamamatsu R7725 photomuliplier tube (PMT) for detection. This PMT is a modified
version with a Pt (platinum) underlay on the photocathode, which was shown to
function at temperatures as low as 2~K~\cite{Meyer2010,Ito2012}.

The PMT was thermally anchored to the 4~K shield of the MSHV
system~\cite{ITO16}. As was done in Refs.~\cite{Ito2012} and \cite{Meyer2010}, the base
circuit for the PMT adopted the split design, where the voltage
dividing resistor chain, which is thermally anchored to the 4~K heat shield,
was separated with a cryogenic ``ribbon cable'' from the charge storing capacitors, located directly on the
PMT. The HV to bias the PMT was supplied using a cryogenic HV coaxial
cable. 

Previous work has shown that the quantum efficiency of the type of PMT used in this experiment decreased by about 10\% from room temperature down to 77~K, but became stable below this point \cite{Meyer2010}.  Such an effect would impact any comparison made between measurements taken far apart in time (many hours to a day) when the photocathode could potentially be at different temperatures, and hence result in different quantum efficiencies between measurements.  To monitor the temperature of the PMT during the experiment, a temperature sensor was attached to the base of the PMT. This part was $<$10~K when the first set of data were acquired.

We chose $^{113}$Sn as our electron source since it provides
monoenergetic electrons, giving a higher sensitivity to the electric
field effect than was possible with an electron source with a
continuous energy spectrum~\cite{GUO12}. The $^{241}$Am source served
as a calibration source and was co-electrodeposited with the $^{113}$Sn in a $~$6.35~mm diameter spot at the center of the high-voltage electrode.  The $^{113}$Sn and $^{241}$Am sources had
activities that corresponded to emission rates of 850~s$^{-1}$ and 195~s$^{-1}$ for 364 to 391-keV electrons and 5.388 to 5.544-MeV $\alpha$ particles, respectively, into the liquid at the time of the experiment; $^{113}$Sn has a half-life of 115 days, whereas the half-life of $^{241}$Am is 432.2 years. 

The gap between the wire mesh ground electrode and the high-voltage
(HV) electrode was 3.8~mm. This is smaller than the range of 364-keV
electrons, which is $\approx 7$~mm.  This arrangement results in a fraction of emitted
electrons hitting the light guide. This
gap size was chosen as a compromise between the following two
considerations: (1) the larger the gap, the larger the fraction of
electrons that ``range out" in LHe in the gap; (2)
the larger the gap, the larger the electrical potential difference that is needed to
achieve the same electric field. With a gap size of 3.8~mm,
$\approx 62$\% of the emitted 364~keV electrons ranged out in
LHe. Those electrons that do not range out in the liquid
in the gap hit the PMMA surface coated with TPB or are backscattered toward the electrode.  Due to the high penetrating power, $<1\%$ of the electron's energy is deposited in the TPB layer, and so the number of photons produced by these electrons directly hitting the surface is negligibly small~\cite{ESTAR}.

One of the HV feedlines of the MSHV system~\cite{ITO16} was used to
supply an electrical potential of up to 15~kV to the HV
electrode. From the HV feedthrough on the CV to the HV electrode, a HV
feedline made of a polytetrafluoroethylene (PTFE) insulated metal wire
was used. When the supplied HV was 15~kV, the electric field in the
gap was 40~kV/cm, and the highest field on the wire mesh was
90~kV/cm.

\subsection{\label{sec:DAQ}Data acquisition system}

The signal from the PMT was sent to an ORTEC 474 timing filter amplifier set with an 100~ns integration time and 100-ns differentiation time.  One of the outputs from the timing filter amplifier was sent to a linear amplifier and discriminator.  The other output was transferred directly to a DDC10 100-MHz waveform digitizer~\cite{SkuTek}.  The discriminator, whose threshold level can be adjusted, was used to trigger the digitizer.  The digitized waveforms were continuously transferred to a DAQ computer and saved to disk as the data were being acquired.

The data for $\alpha$ particles and electrons were acquired simultaneously with the trigger threshold set to a level that corresponded to $\approx 3$ photoelectrons (PEs).  This trigger level was chosen to avoid triggering on single photoelectron pulses that are generated by PMT dark current emission or afterpulses generated in the liquid.  In terms of the number of photoelectrons, this threshold drifted slightly over the duration of the experiment due to a drift in the PMT gain.  The gain drift was a result of a lag in the cooling rate of the internal PMT dynode structure relative to that of the glass phototube.  To account for these drifts, the signals within a given dataset were calibrated against the single photoelectron response of the PMT for that same dataset.  This is further discussed in Sec.~\ref{sec:apspectrum}.  

\begin{figure}[]
	\includegraphics[width=\columnwidth]{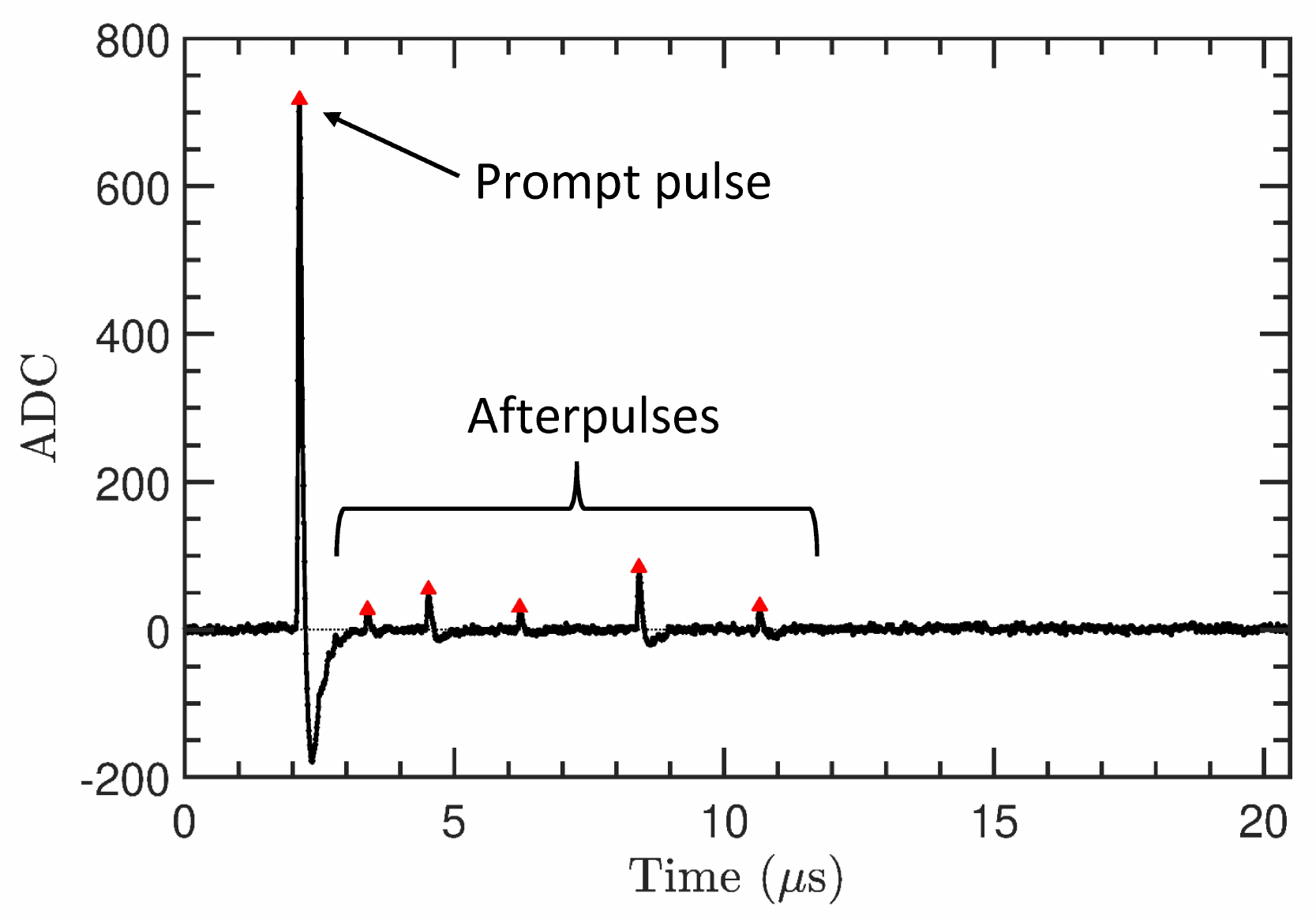}
	\caption{A sample digitized waveform.  The red triangles mark the locations of the detected peaks.}
	\label{fig:waveform}
\end{figure}

For each trigger, a 20.47-$\mu$s waveform was captured.  Each waveform consisted of 2048 samples which were captured at rate of 10~ns per sample.  The trigger was offset by 2.1~$\mu$s from the start of the waveform time window so that there was 2.1~$\mu$s of pretrigger data and 18.37~$\mu$s of posttrigger data.  Figure~\ref{fig:waveform} shows a sample digitized waveform.

It was found after the data had already been acquired that input reflection at the linear amplifier used in the data acquisition system caused a reduction in the amplitudes of the digitized signals.  To correct for this, a pulse generator was used to measure the reduction as a function of the amplitude of the input signal.  We found that only the large signals produced by $\alpha$ scintillation were affected, and the amplitudes of these signals were corrected using the calibration measurements made with the pulse generator.

\section{\label{sec:data}Data and analysis}

\subsection{\label{sec:meas}Range of the measurements}

Data were acquired for six different temperatures, ranging from 0.44~K to 3.12~K, at a pressure of approximately 600~Torr rather than at saturated vapor pressure (SVP).  Measurements were made at two temperatures above the $\lambda$ transition and four below it.  For a set of measurements at a fixed temperature, the potential difference between the electrode and the wire mesh that acts as a ground plane was ramped up from 0 to 15~kV.  At 0.44~K, measurements were made with both polarities on the high voltage electrode.  Table~\ref{tab:datasets} summarizes the parameters for each of the datasets.  The temperatures and pressures shown represent the average of the start and end values for each dataset.  Here, we are referring to a dataset as a complete series of measurements over the range of voltages -- for both polarities, when applicable -- at one temperature.  

\begin{table}[!b]
	\caption{\label{tab:datasets} Dataset and conditions}
	\begin{ruledtabular}
		\begin{tabular}{ccccc}
			\textrm{Dataset} &
			\textrm{$T$ [K]}&
			\textrm{$P$ [Torr]}&
			\textrm{$E$ [kV/cm]}&
			\textrm{$\rho$ [g/cm$^3$]}\\
			\colrule
			\RomanNumeralCaps{1} &	0.44 & 607 & (-40, 40) & 0.1466  \\ 
			\RomanNumeralCaps{2} &	0.84 & 611 & (0, 40) & 0.1466  \\ 
			\RomanNumeralCaps{3} &	1.15 & 600 & (0, 40)   & 0.1466  \\ 
			\RomanNumeralCaps{4} & 	1.65 & 597 & (0, 40)  & 0.1468 \\ 
			\RomanNumeralCaps{5} &	2.35 & 601 & (0, 40)  & 0.1472 \\
			\RomanNumeralCaps{6} &	3.12 & 627 & (0, 40) & 0.1418 \\ 
		\end{tabular}
	\end{ruledtabular}
\end{table}

Uncalibrated ruthenium oxide (ROX) sensors were used to monitor the temperature of the experimental volume.  The ROX sensors have a stated accuracy of $\pm25$ mK at 0.5~K and $\pm75$~mK at 2.0~K~\cite{Lakeshore}.  The largest temperature variation ($2.43 - 2.27$~K), one which was well outside the stated uncertainty of the ROX sensors was exhibited by dataset \RomanNumeralCaps{5}, and was the result of difficulty in stabilizing the temperature over the duration of these measurements.  With the exception of the 3.12~K dataset, where the pressure varied between 648 and 605~Torr between the start and end of the measurement cycle, the variation in pressure for all other datasets was only a few Torr during the data acquisition of that dataset.

In total, the measurements consisted of 43 subsets of data, and for each subset at a particular temperature, pressure, and electrode voltage setting, $2\times10^{5}$ events were acquired.  Of these, approximately $30\%$ were $^{241}$Am $\alpha$ particles [5485.7~keV (85.2\%), 5443.0~keV (12.80\%), 5388.0~keV (1.40\%)] and $70\%$ were $^{113}$Sn conversion electrons [363.76~keV (28.2\%), 387.46~keV (5.48\%), and 391.0~keV (1.245\%)].  Here the paired values in square brackets represent the energies of the particles and their corresponding branching ratios, restricted to cases where the branching ratio is at least 1\%.  The average energy, $\overbar{\varepsilon}$, for the $\alpha$ particles and electrons were 5446 and 368~keV, respectively. An event trigger was categorized as either an $\alpha$ particle or an electron by the number of photoelectrons in the prompt pulse.  The total acquisition time for each data subset was approximately 10 minutes, and each subset was calibrated with the single photoelectron distribution acquired in the same subset.

\begin{figure*}[]
	\includegraphics[width=\textwidth]{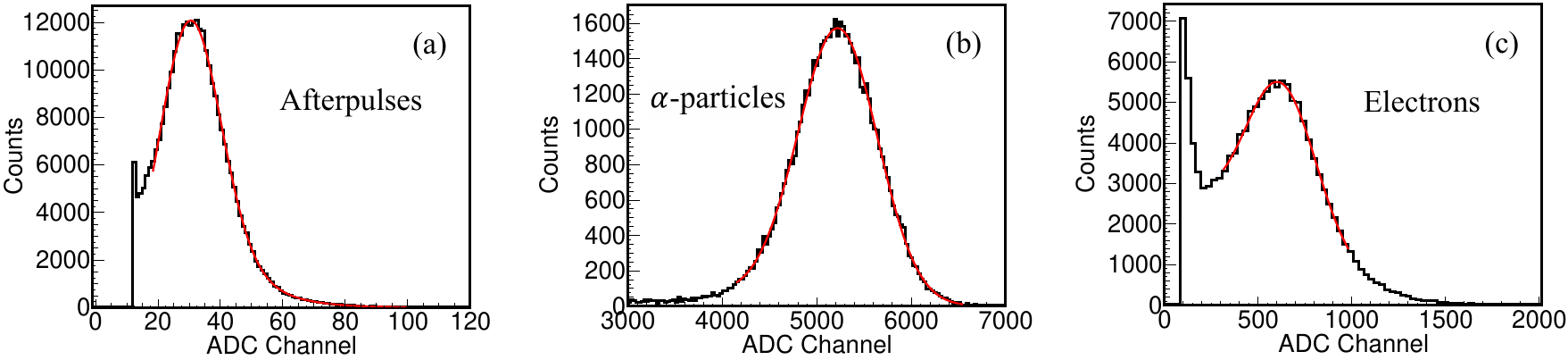}
	\caption{(a) A sample fit of the afterpulse ADC spectrum for one subset of the data.  The fit returns the PMT gain, $Q_{1}$, which is the ADC channel corresponding to one photoelectron.  The left edge corresponds to an analysis threshold for identifying pulses above the noise level. (b) A fit of the $\alpha$-particle ADC spectrum for one subset of the data using the fit function given by Eq.~\ref{eq:alphafitfunction}.  (c) A fit of the electron ADC spectrum for one subset of the data using the fitting function given by Eq.~\ref{eq:electronfitfunction}.}
	\label{fig:spectrafits}
\end{figure*}

\subsection{\label{sec:waveformanalysis}Waveform analysis}

A peak detection algorithm is applied to each digitized waveform.  The algorithm detects the amplitude and location of each peak in the waveform. Examples, marked by red triangles, are shown in Fig.~\ref{fig:waveform}.  Peaks are classified based on their time locations as one of the following:  a pretrigger peak, a trigger peak (prompt pulse), an afterpulse.  The afterpulses are generated by single photoelectrons, and their distribution is used to determine the PMT gain.  For this purpose, only afterpulses after 6~$\mu$s are used for fitting the afterpulse ADC spectrum.  Furthermore, to prevent distortion to the single photoelectron spectrum from overlapping afterpulses, a pulse time separation analysis cut is made to remove afterpulses separated by less than 800~ns.  However, these analysis cuts are not used in analyzing the time spectrum of the afterpulses, results of which will be presented in a forthcoming paper.

\subsection{\label{sec:fitting}Spectrum fitting}

\subsubsection{\label{sec:apspectrum}Afterpulse spectrum}

Single photoelectron (SPE) pulses in the afterpulse region of both the $\alpha$ and electron triggered waveforms are used to characterize the PMT gain, providing for the conversion between the measured ADC channel to number of photoelectrons. The SPE spectrum is fitted with the PMT response function proposed by Bellamy~\textit{et al.}~\cite{Bellamy1994}.  There are seven free parameters in the fitting function: $Q_{0}$, $\sigma_{0}$, $Q_{1}$, $\sigma_{1}$, $\mu$, $w$, and $\alpha$.  The parameters $Q_{0}$ and $\sigma_{0}$ characterize the mean and width of the pedestal distribution.  $Q_{1}$, $\sigma_{1}$, and $\mu$ characterize the PMT gain, distribution width, and source intensity, respectively.  The final two parameters, $w$ and $\alpha$, describe the discreet background distribution.  Once $Q_{0}$ and $\sigma_{0}$ are determined from a fit of the pedestal distribution, their values are fixed in the fit of the SPE distribution. Shown in Fig.~\ref{fig:spectrafits}(a) is a fit of the SPE distribution for one subset of the data.  The energy resolution of the SPE response ($\sigma_{1}/Q_{1}$) obtained from the fit is $\approx 32$\% and is consistent with the value expected for the type of PMT used in this experiment.  The fitting procedure is performed to determine the PMT gain for each data subset.

\subsubsection{\label{sec:alphaspectrum}$\alpha$ prompt scintillation spectrum}

The $\alpha$ ADC spectrum of the prompt scintillation signal is fitted with an analytic peak-shape function proposed by Bortels~and~Collaers~\cite{Bortels1987} to fit $\alpha$ spectra in Si detectors.  The function consists of the convolution of a Gaussian with the weighted sum of two left sided exponential functions to model the low-energy tail in the spectrum and is given by
\begin{IEEEeqnarray}{rCl}
	f(x) & = & \frac{A}{2} \Biggl\{ \frac{1-\eta}{\tau_{1}} \exp\left(\frac{x-\mu}{\tau_{1}}  + \frac{\sigma^{2}}{2\tau_{1}^{2}} \right) 
	\nonumber\\
	&&    \times   \erfc\left[ \frac{1}{\sqrt2} \left( \frac{x-\mu}{\sigma}   + \frac{\sigma}{\tau_{1}}     \right) \right]  
	\nonumber\\
	&&  + \frac{\eta}{\tau_{2}} \exp\left(\frac{x-\mu}{\tau_{2}}  + \frac{\sigma^{2}}{2\tau_{2}^{2}} \right)
	\nonumber\\
	&&   \times  \erfc\left[ \frac{1}{\sqrt2} \left( \frac{x-\mu}{\sigma}  + \frac{\sigma}{\tau_{2}}  \right) \right]   \Biggr\}.  
	\label{eq:alphafitfunction}
\end{IEEEeqnarray}
In Eq.~\ref{eq:alphafitfunction}, $\mu$ and $\sigma$ are the mean and standard deviation of the Gaussian component, $\tau_{1}$ and $\tau_{2}$ are the parameters of the two exponential functions, $\eta$ is a weighing factor, and $A$ is the overall normalization.  A third exponential can also be included to further improve the fit in some instances, but we find that two terms are sufficient for a good fit to our data and choose not to include the additional term in our analysis.  In principle, the $\alpha$ spectrum for $^{241}$Am is fitted with multiple peaks and Eq.~\ref{eq:alphafitfunction} is written as a sum over the number of peaks in the fit.  However, we choose to fit the spectrum with only one peak because the secondary peaks in the spectrum are very close in energy to the primary peak and also have much smaller branching ratios.  The number of photoelectrons in the peak, $\overbar{N}_{PE}$, is equal to $\mu/Q_{1}$, where $Q_{1}$ is the ADC gain determined from a fit of the afterpulse ADC spectrum.  Figure~\ref{fig:spectrafits}(b) shows a fit to the $\alpha$ prompt scintillation spectrum of one data subset using Eq.~\ref{eq:alphafitfunction}.

\subsubsection{\label{sec:electronspectrum}Electron prompt scintillation spectrum}

\begin{figure}[htb!]
	\includegraphics[width=\columnwidth]{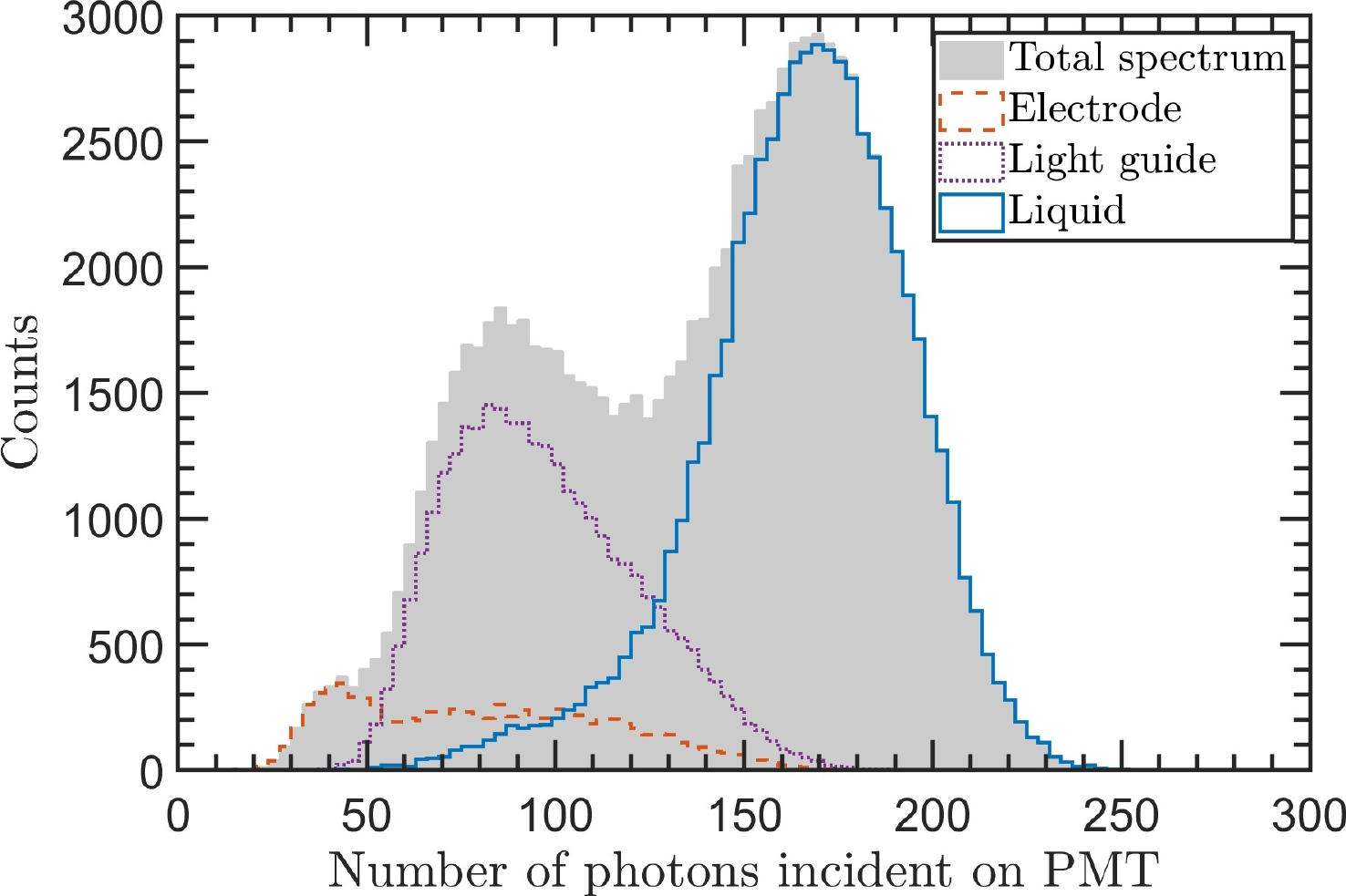}
	\caption{Simulated spectrum of the scintillation signal produced by 364-keV electrons in the experimental setup.  The contributions to the total spectrum (solid-filled gray) are from back-scattered electrons that range out in the electrode (dashed red), electrons that range out in the PMMA light guide (dotted purple), and electrons that range out in LHe (solid blue).  The PMT quantum efficiency, resolution, and noise are not included in the simulation. }
	\label{fig:simspectrum}
\end{figure}

The $^{113}$Sn electron prompt scintillation spectrum is fitted by a function composed of the sum of three Gaussians, each representing one of the energies of the conversion electrons, and an exponential function.  The fit function is given by
\begin{IEEEeqnarray}{rCl}
	f(x) & = & N \Biggl\{  a_{1}\exp \left[ -\frac{1}{2}\left( \frac{x - \mu}{\sigma}   \right)^2 \right]
	\nonumber\\
	&&   +  a_{2}\exp \left[ -\frac{1}{2}\left( \frac{x - b_{2}\mu}{\sigma}   \right)^2 \right] 
	\nonumber\\
	&&  + a_{3}\exp\left[ -\frac{1}{2}\left( \frac{x - b_{3}\mu}{\sigma}   \right)^2 \right]  \Biggr\}
	\nonumber\\
	&&  +  A\exp\left( -\lambda x \right).   
	\label{eq:electronfitfunction}   
\end{IEEEeqnarray}
There are five free parameters in the fit.  Two of the parameters are the mean, $\mu$, and width, $\sigma$, of the Gaussian function for the 364-keV peak.  The mean values of the other two peaks are expressed as the mean of the first peak weighted by the energy of these two peaks relative to the first.  The widths are taken to be the same for all three peaks, and the amplitude of each peak is weighted by its branching ratio.  $N$ is the overall normalization for the Gaussian component.  Two other parameters, $A$ and $\lambda$, characterize the exponential component.

A fit of the electron prompt scintillation spectrum of one subset of the data is shown in Fig.~\ref{fig:spectrafits}(c), where the left edge is due to an analysis threshold.  The events that do not range out in the gap between electrodes form a low-energy tail that extends below this threshold as shown in the simulated spectrum in Fig.~\ref{fig:simspectrum}, which is obtained with GEANT4~\cite{GEANT4} and the PENELOPE Low Energy Electromagnetic Physics Model~\cite{PENELOPE1, PENELOPE2}.  In the simulation, the geometry of the light detection setup (Fig.~\ref{fig:setup}) is modeled in GEANT4.  Electrons with an energy of 364~keV are emitted from the surface of the electrode isotropically into the liquid, generating EUV photons, some of which strike the TPB surface.  These photons are then wavelength shifted by the TPB, and a fraction of them are captured and transmitted down the light guide,  through the sapphire window, and finally end up at the PMT where they are counted.  The PMT quantum efficiency, resolution, and noise are not included in the simulation.  There are two components that contribute to the low energy tail, one from electrons that range out in the PMMA light guide ($\approx 30$\%), and the second from electrons that are back-scattered and range out in the high-voltage electrode ($\approx 8$\%).  Their contribution, however, is smaller than the fraction of electrons that range out in the liquid ($\approx 62$\%), which forms a distinct peak in the spectrum.

\subsection{\label{sec:results}Results}

\subsubsection{\label{sec:pmtgain}PMT gain}

\begin{figure}[]
	\includegraphics[width=\columnwidth]{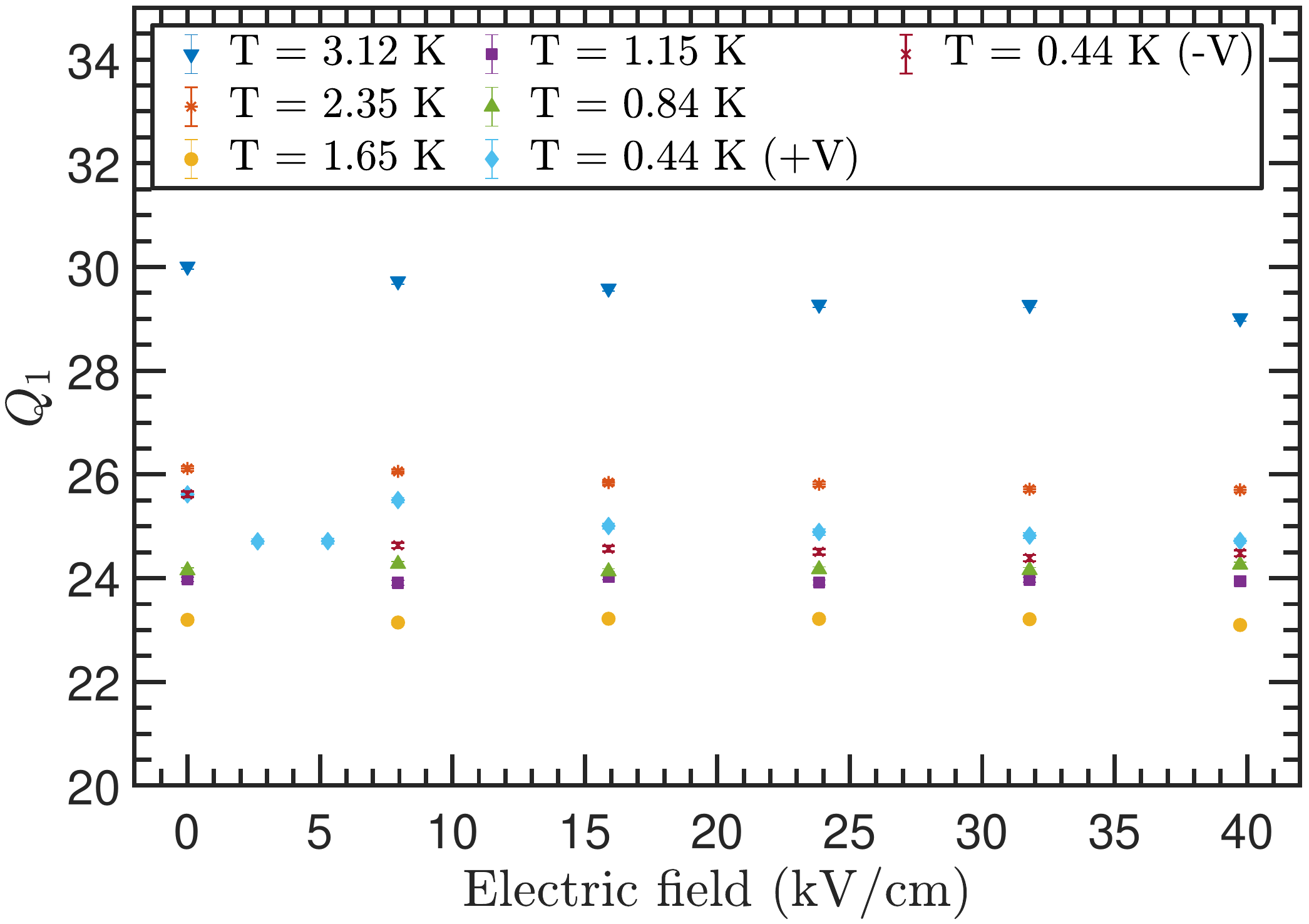}
	\caption{PMT gain from fits of the afterpulse spectra for all 43 datasets acquired during this experiment.  The size of the statistical error bars is smaller than the data points.}
	\label{fig:PMTgain}
\end{figure}

The PMT gain for all subsets of the data acquired in the experiment is shown in Fig.~\ref{fig:PMTgain}.  The time order of the data is as follows: 3.12~K (\RomanNumeralCaps{6}), 2.35~K (\RomanNumeralCaps{5}), 0.44~K (\RomanNumeralCaps{1}), 0.84~K (\RomanNumeralCaps{2}), 1.15~K (\RomanNumeralCaps{3}), and 1.65~K (\RomanNumeralCaps{4}).  For the 0.44~K data, the order of the voltage settings is the following:  0~kV, 3~kV, 6~kV, 9~kV, 12~kV, 15~kV, 1~kV, 2~kV, -3~kV, -6~kV, -9~kV, -12~kV, -15~kV.  We observed a trend of decreasing PMT gain with time, highlighting the importance of individually calibrating each subset of the data with the afterpulse distribution from the same subset.

\subsubsection{\label{sec:promptpes}Mean number of prompt photoelectrons}

\begin{figure*}
	\includegraphics[width=\textwidth]{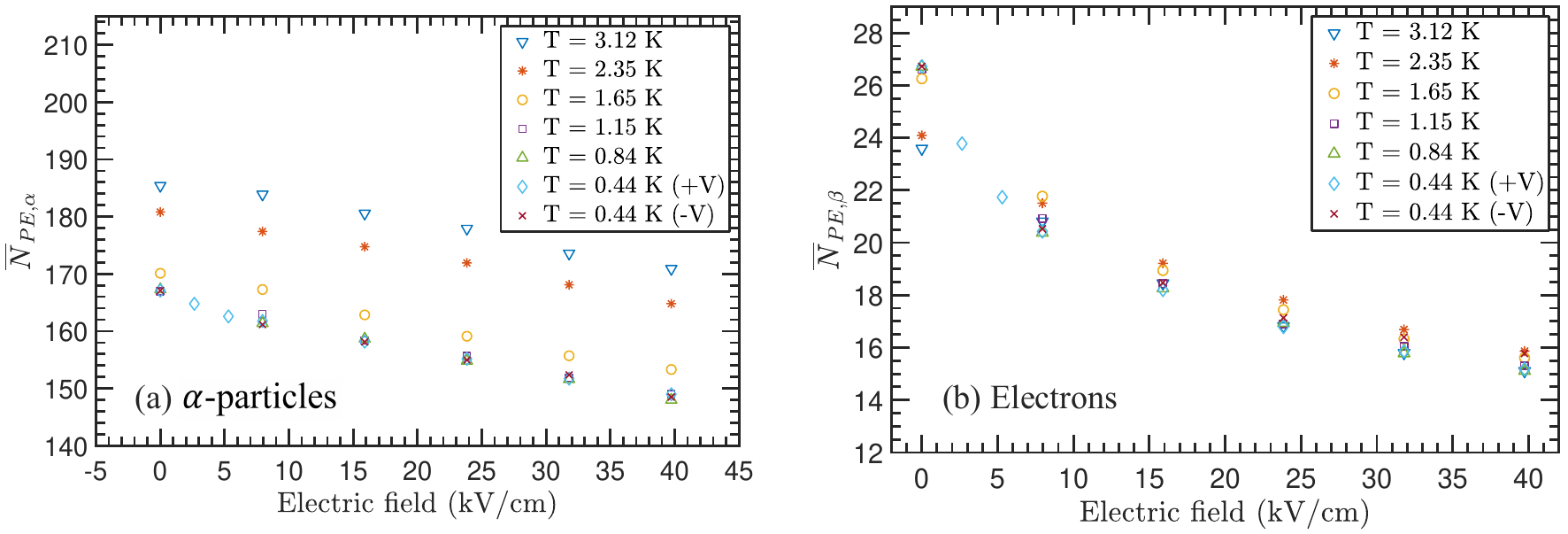}
	\caption{The absolute mean number of detected PEs in the prompt signal (first 100~ns) as a function of the applied electric field for (a) $^{241}$Am $\alpha$ particles and (b) 364~keV electrons from $^{113}$Sn.}
	\label{fig:abslightyield}
\end{figure*}

The mean number of photoelectrons in the prompt signal (first 100~ns), $\overbar{N}_{PE}$, is defined as the fitted ADC channel peak value of the $\alpha$(electron) spectrum divided by the PMT gain, $Q_{1}$, measured at the same temperature and field.  The mean number of prompt photoelectrons as a function of electric field for all temperature datasets is shown in Figs.~\ref{fig:abslightyield}(a) and \ref{fig:abslightyield}(b) for $\alpha$ particles and electrons, respectively.

\begin{figure*}[]
	\includegraphics[width=\textwidth]{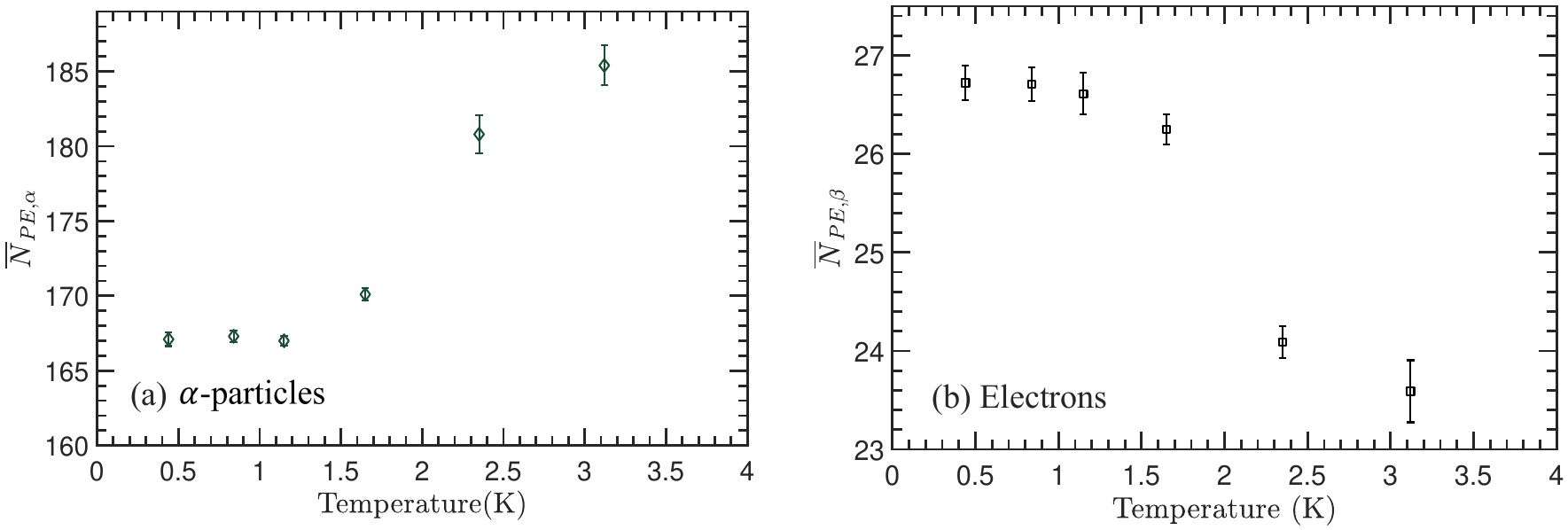}
	\caption{The absolute mean number of detected PEs in the prompt signal (first 100~ns) at zero field as a function of temperature for (a) $^{241}$Am $\alpha$ particles and (b) 364~keV electrons from $^{113}$Sn.}
	\label{fig:zerofieldPE}
\end{figure*}

For the zero-field $\alpha$ data in this work, an approximately 8\% decrease in the detected prompt scintillation yield between 2.35~K and 0.44~K is observed (Fig.~\ref{fig:zerofieldPE}(a)).  In comparison, the data from Ito~\textit{et al.}~\cite{Ito2012} which were taken at SVP show about a 9\% reduction over approximately the same temperature range.  Density effects are a likely explanation for this small difference.  For more discussion, see Sec.~\ref{sec:discussalphafielddepend}. 

For the zero-field electron data, we observe an increase in the detected prompt scintillation signal with decreasing temperature as shown in Fig.~\ref{fig:zerofieldPE}(b).  The trend exhibited by these data is the inverse of the trend observed for $\alpha$ particles (Fig.~\ref{fig:zerofieldPE}(a)).  In comparison, the results for electron scintillation from Kane~\textit{et al.}~\cite{Kane1963} are strikingly different.  They observe a relatively flat yield for temperatures above the $\lambda$ transition and a similarly flat yield for temperatures below it.  However, the low-temperature yield is reduced by about 5\% relative to their high-temperature yield and a very steep, almost discontinuity-like change around the $\lambda$ transition is observed, forming a step like function.  We discuss in detail the origin of the observed temperature dependence in Sec.~\ref{sec:electronfielddepend}. Our data and our understanding of the phenomenon are not consistent with the sharp discontinuity observed by Kane~\textit{et al.}. Their observation could be the result of their particular methodology and experimental setup.

\subsubsection{\label{sec:lightyield}Normalized prompt scintillation yield}

\begin{figure*}[]
	\includegraphics[width=\textwidth]{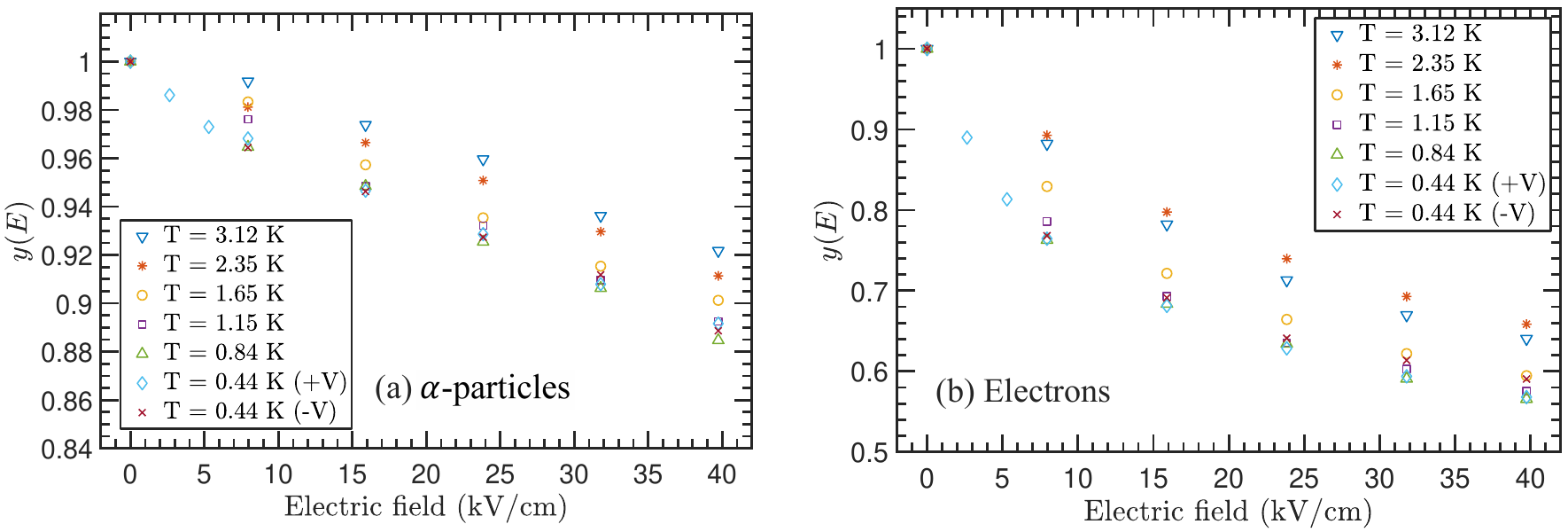}
	\caption{The normalized detected scintillation yield of the prompt signal (first 100~ns) as a function of the applied electric field at six different temperatures for (a) $^{241}$Am $\alpha$ particles and (b) 364~keV electrons from $^{113}$Sn.  The temperature dependence seen in the electron data is due to the signal integration time and is discussed further in Sec.~\ref{sec:electronfielddepend}.}
	\label{fig:normyield}
\end{figure*}

For analyzing the effect of the applied field on the scintillation signal, it is more convenient to examine the zero-field normalized scintillation yield, $y$, which is defined as  
\begin{equation}\label{eq:normyielddefine}
y(E,T) \equiv \frac{\overbar{N}_{PE}(T, E)}{\overbar{N}_{PE}(T, E = 0)}.
\end{equation}
Here, $\overbar{N}_{PE}$ is mean number of detected prompt photoelectrons at a given temperature $T$ and electric field $E$ and $\overbar{N}_{PE}(T, E = 0)$ is the number of photoelectrons detected at the same temperature $T$ and zero field.  In Fig.~\ref{fig:normyield}, the zero-field normalized detected prompt scintillation yield as a function of the applied electric field is shown.  The normalized $\alpha$ prompt scintillation yield exhibits an interesting, and perhaps surprising, temperature dependence of the yield reduction with field.  This feature was not observed by Ito~\textit{et al.}~\cite{Ito2012} over the temperature range of 0.2~K to 1.1~K.  Consistent with their observation is the absence of a temperature dependence in our data below 1.15~K.  The yield reduction that we observe between 0 and 40~kV/cm at 0.44~K is about 11\%, which is in good agreement with the results from Ito~\textit{et al.}.  Small differences are attributable to the uncertainty in the electric field in both experiments.  For this experiment, the estimated uncertainty in the gap spanned by the high voltage electrode and the ground grid, and hence the electric field, is $\approx 5-10$\%.  Furthermore, the measurements in this work are acquired at $\approx 600$~Torr while those from Ref.~\cite{Ito2012} are taken at SVP, and density effects may play a role.  The electron light yield and its field dependence will be discussed in more detail in Sec.~\ref{sec:electronfielddepend}.

\subsubsection{\label{sec:polarityeffect}Effect of voltage polarity on electrons}

\begin{figure}[]
	\includegraphics[width=\columnwidth]{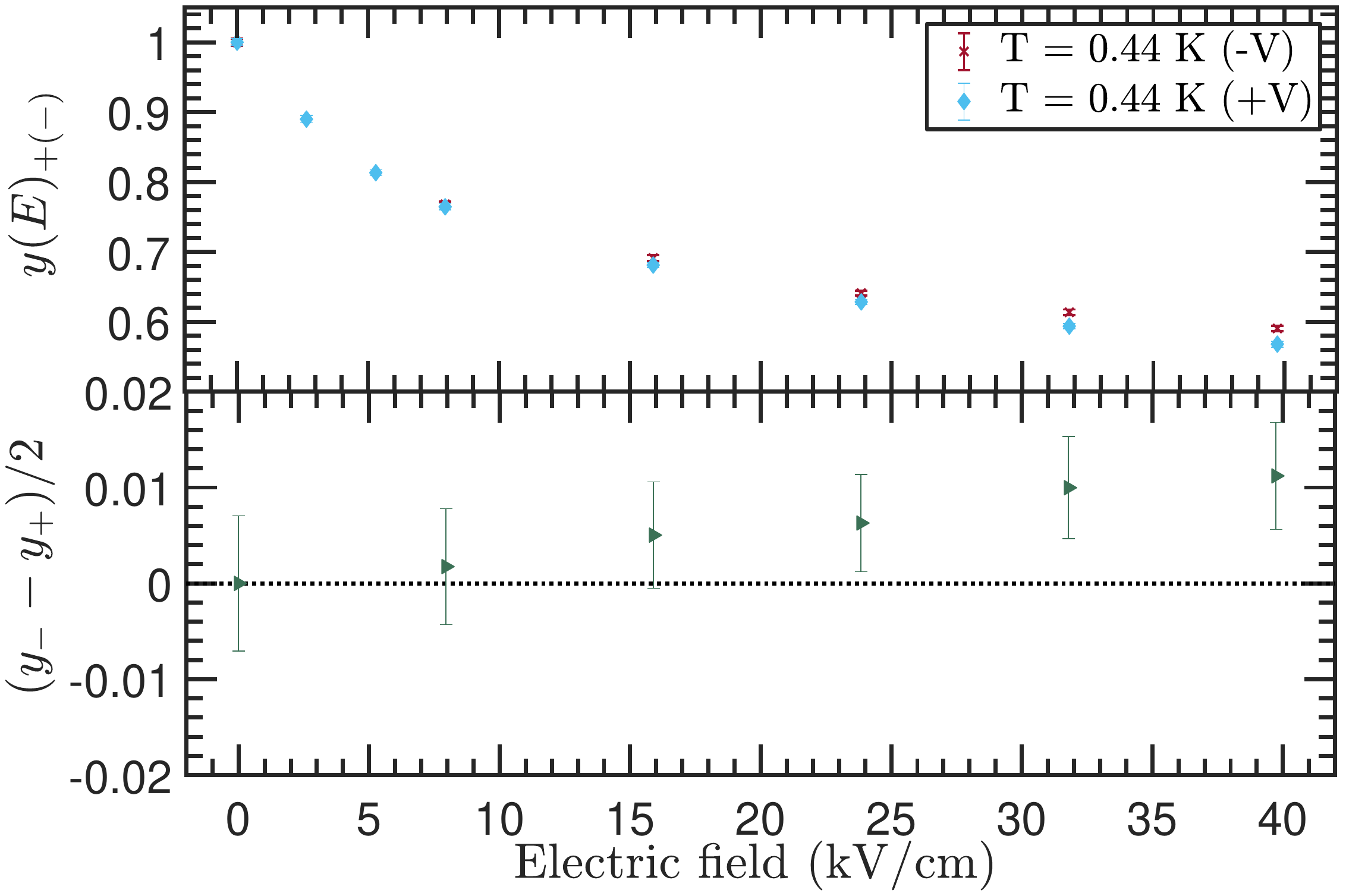}
	\caption{Top: The normalized scintillation yield in the prompt signal as a function of electric field for the two voltage polarities at $T = 0.44$~K. Bottom: The half difference in the yield between the polarity measurements.}
	\label{fig:polarityeffect}
\end{figure}

Given that most of the $\approx 364$~keV electrons from the radiation source used in this experiment will transverse the entire high field region between the electrodes, the effect of the electric field on their energy and trajectories must be considered.  For this purpose, measurements were obtained for a potential difference between the electrode and wire mesh of -15~kV to +15~kV.  These measurements are only made at $T = 0.44$~K, and the normalized scintillation yield of the prompt signal at this temperature for the two electrode polarities is shown in Fig.~\ref{fig:polarityeffect} (top plot).  Note that the horizontal axis represents the absolute value of the electric field.  

With a range of $\approx 7$~mm for $^{113}$Sn conversion electrons in LHe and a high field region gap size of $\approx 3.8$~mm, only electrons tracks with steep angles relative to the electrode surface normal will range out in the liquid.  This implies that the effect of the electric field (relative to no field) should be of order a few percentages (40~kV/cm~$\times$~0.38~cm/364~kV = 4\%).  Plotted in Fig.~\ref{fig:polarityeffect}(bottom plot) is the half-difference between the negative and positive polarity light yield, showing that the effect is indeed small and only $\approx 1$\% at the highest field measured.  Also notice that the normalized detected light yield for the negative polarity data is higher than that for the positive polarity data over the entire range of electric fields.  This is consistent with the expectation that the electron energy is slightly boosted by the field when the voltage on the source plated electrode is negative.

\subsubsection{\label{sec:alphabetaratio}$\alpha/\beta$ ratio}

The co-electrodeposition of the $\alpha$ and conversion electron sources in our experiment allows for simultaneous measurements of the scintillation yield produced by these two sources.  From this, a determination of the $\alpha/\beta$ ratio, or ``quenching factor," $\eta$, is quite straightforward.  This quantity is the ratio of the number of prompt photoelectrons from $\alpha$ excitation per unit energy deposition to prompt photoelectrons from electron excitation per unit energy deposition and is defined as
\begin{equation}
\eta \equiv \frac{R_{\alpha}}{R_{\beta}}, 
\end{equation}
with 
\begin{equation}\label{eq:alphabetaratio}
R_{\alpha(\beta)}= \frac{\mbox{ $\overbar{N}_{PEs, \alpha(\beta)}$}}{\mbox{$\varepsilon_{\alpha(\beta)}$} \times \mbox{$\omega_{\alpha(\beta)}$}}.
\end{equation}
The parameters $\varepsilon$ and $\omega$ are the primary particle energy and the scintillation photon geometric acceptance factor, respectively.  The latter is the fraction of emitted EUV photons generated by the corresponding particle that strike the TPB surface, which is determined through a GEANT4 simulation using the PENELOPE Low Energy Electromagnetic Physics Model.  For every $\alpha$ decay, $N_{\gamma, \alpha}$ primary photons are generated and emitted isotropically, and of those, the number that strike the TPB surface on top of the light guide, $N_{\text{TPB},\alpha}$, is counted.  A histogram of $N_{\text{TPB},\alpha}$ is accumulated and fitted with a Gaussian.  The fitted mean value is divided by the number of primary photons to obtain the photon geometric acceptance factor, $\omega_{\alpha}$.  The same procedure is carried out for electrons from $^{113}$Sn to obtain the electron geometric acceptance factor, $\omega_{\beta}$, but the fit function used for the electrons is given by Eq.~\ref{eq:electronfitfunction}.  The simulation is carried out for each liquid density listed in Table~\ref{tab:datasets}.  For $\alpha$ particles, the average geometric acceptance factor is $0.350 \pm 0.001$ with negligible changes with liquid density.  The factor for electrons is $0.378 \pm 0.002$ between 0.44 and 1.15~K, $0.379 \pm 0.002$ between 1.65 and 2.35~K, and $0.377 \pm 0.003$ at 3.12~K.  All stated errors are statistical.  We note that the estimated uncertainty in the positions of the components of the light collection system is $\approx 5-10$\%.  However, the systematic errors on the absolute geometric acceptances due to this uncertainty is small ($\approx 1$\%) and do not affect any of the discussion that follows.

\begin{figure}[]
	\includegraphics[width=\columnwidth]{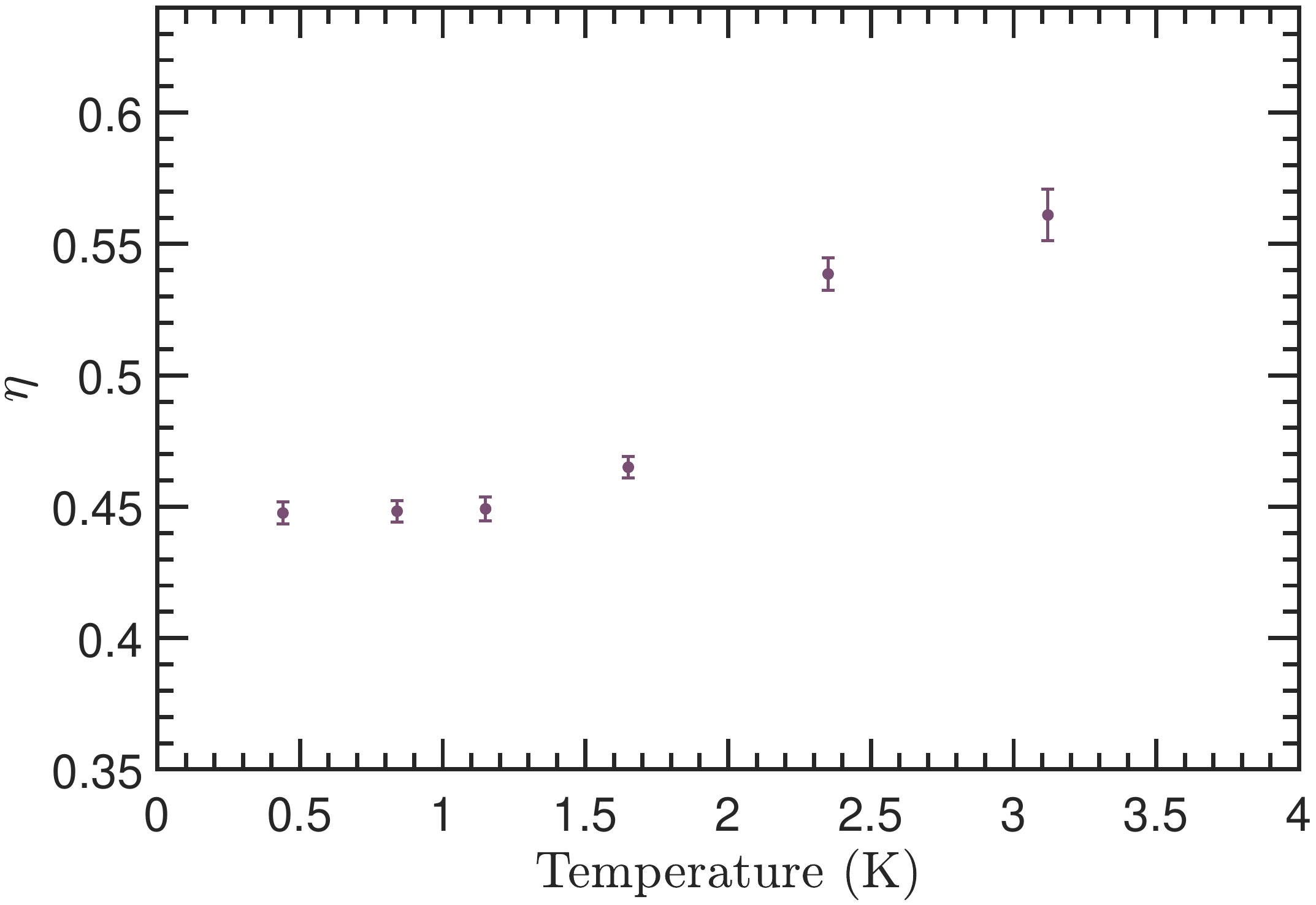}
	\caption{The $\alpha/\beta$ ratio as a function of temperature corrected for geometric acceptance. Note that the prompt signals from both particles are integrated over the first 100~ns and the apparent temperature dependence of the ratio is a result of this.  Refer to Sec.~\ref{sec:electronfielddepend} for more details.}
	\label{fig:alphabetaratio}
\end{figure}

Figure~\ref{fig:alphabetaratio} shows the ratio, $\eta$, as a function of the temperature at zero field.  The error bars are statistical and are derived from the errors in the fits of the spectral peaks.  The ratio has a striking temperature dependence, and this dependence is not merely a density effect because simulations shows that the geometric acceptance factor changes very little over the temperature range in the experiment for both particle types.  This effect is tied to the temperature dependence of the zero-field scintillation yield as shown in Fig.~\ref{fig:zerofieldPE}.  The reason behind this is discussed in more detail in Sec.~\ref{sec:discussion}.

Several previous investigations of the $\alpha/\beta$ ratio have been made.  One of the earliest measurements was made by Miller\cite{Miller1964}, who measured a value of 0.182 (3\% stated uncertainty).  The two sources used in his experiment were $^{241}$Am and $^{60}$Co, and the wavelength shifters were POPOP and DPS.  Presumably, the measurement was made at $\approx 4$~K, although the exact temperature is not stated in the paper.  Moreover, the integration time of the signal pulse is not specified so it is not known whether the ratio is only of the prompt component of the scintillation or has a contribution from the afterpulses as well.  Given these uncertainties, a direct comparison of the results from this work with those of Miller is tenuous.  Nevertheless, at face value, the results in this work do not appear to be reconcilable with those of Miller, and the reason behind the disagreement is unclear.

Other measurements of the $\alpha/\beta$ ratio in the literature include those from Adams~\cite{AdamsThesis} and Adams~\textit{et al.}~\cite{Adams1998}.  In the former work, it is stated that 35\% of an electron's energy goes into scintillation light while for an $\alpha$ particle it is only 10\%.  This implies an $\alpha/\beta$ ratio of 0.29.  However, in Ref.~\cite{Adams1998}, the fraction of energy that goes into scintillation is stated to be 24\% and 10\% for the electron and $\alpha$, respectively.  The ratio implied by these values is 0.42, and this value appears to be consistent with our lowest temperature result of 0.45.
  
McKinsey~\textit{et al.}\cite{McKinsey2004} has also measured the ratio, stating a value of $0.50\pm 0.10$.  The $\alpha$ and electron sources used in their experiment were $^{241}$Am and $^{113}$Sn, respectively, and the wavelength shifter used was TPB doped polystyrene at 40\% concentration; thus the setup of their experiment is very similar to the one used in this work.  Within the stated uncertainty, their result is consistent with our new results.

\section{\label{sec:discussion}Discussion}

\subsection{\label{sec:discussalphafielddepend}Temperature and electric field dependence of $\alpha$ prompt scintillation }

The temperature dependence of the prompt scintillation yield at zero
electric field for $\alpha$ particles plotted in Fig.~\ref{fig:zerofieldPE}(a) is consistent with what
was reported in Ref.~\cite{Ito2012}. A similar temperature dependence,
that is, a reduction in scintillation yield with decreasing
temperature, was observed in the past~\cite{Mos63,Hereford1966,Mel69,Man71,Rob71,Rob72,Rob73}. 
As pointed out in Ref.~\cite{Ito2012}, however, the results from these past experiments
cannot be directly compared to the results reported here and in
Ref.~\cite{Ito2012}. The electronics integration time for the
scintillation pulse was $\approx 1$~$\mu$s in the previous experiments,
whereas it was $\approx 100$~ns in the experiments reported here and in
Ref.~\cite{Ito2012}. The scintillation pulse in the past experiments are
likely to have included part of what we call ``afterpulses,'' which
have their own temperature dependence~\cite{Ito2012}.

In a series of papers, Hereford and collaborators (see Ref.~\cite{Rob73} and references therein) described a model for scintillation
light production. They attributed LHe scintillation to radiative
destruction of some metastable states due to interactions with some
collision partners. The temperature dependence of the scintillation
yield comes from the temperature-dependent diffusion constant of these
species in LHe, which affects the rate of expansion of the column that
contains these species.

However, this picture is incompatible with our current understanding
of the phenomenon. Here, the prompt scintillation is due to
radiative decay of excited singlet species (excimers and atoms). 
Furthermore, the temperature inside the column created by the passage 
of an $\alpha$ particle is about 2~K irrespective of the temperature of the 
bulk LHe when it is below $\approx 1.8$~K~\cite{Ito2012}.

The temperature dependence of the zero-field scintillation yield can
be more naturally attributed to the effect of phonons and rotons
confining the columns of positive and negative species. At lower
temperatures, where the density of phonons and rotons is lower, the
column expands faster, lowering the recombination rate and reducing
the scintillation light emitted during the electronics integration
time.

The same process can explain the temperature dependence of the effect
of an electric field on the yield of the $\alpha$-induced scintillation 
shown in Fig.~\ref{fig:normyield}(a). As discussed in Ref.~\cite{Ito2012}, 
in the presence of an electric field, the fraction of ions that recombine 
depends on a single parameter $f=\sqrt{\pi} \epsilon_0 b E/(N_0 e)$, where 
$b$ is the Gaussian width of the charge column and $N_{0}$ is the number of charges 
per unit length along the track. At lower temperatures, $b$ tends to expand 
faster, but an increase in $b$ has the same effect as increasing the electric 
field, $E$.  Thus, at a given field, its effect is magnified when $b$ increases.

\subsection{\label{sec:electronfielddepend}Temperature and electric field dependence of electron prompt scintillation }

At the highest field measured, the electron data show a reduction in the prompt scintillation yield of $>30$\%.  Furthermore, the normalized prompt scintillation yield as a function of applied field in Fig.~\ref{fig:normyield}(b) possesses an intriguing feature -- a conspicuous temperature dependence that is most notable when comparing data obtained at temperatures above the $\lambda$ transition against those acquired below it.  Below we discuss these observed features in more detail and outline a possible explanation for our observations.

\subsubsection{\label{sec:model}Model of scintillation yield vs electric field}

Consider the following model for the dependence of the prompt scintillation yield on electric field.  Let $a$ be the number of electrons and positive ions that recombine as singlets, $b$ the number that recombine as triplets, and $x$ the ratio of the number of singlet excitations to the total number of ionizations.  Then the prompt scintillation yield as a function of field is 
\begin{equation}
Y(E) = a - \frac{a}{a + b}I(E) + x(a + b),
\end{equation}
with $I(E)$ being proportional to the ionization current. The first term corresponds to the prompt scintillation yield due to excimers that form from recombination of ionization in the absence of an electric field. The second term represents the reduction due to the electric field.  Relating the fraction $x$ to the total number of ionizations, the third term represents the prompt contribution from directly excited atoms. Normalizing the current to the value at $E = \infty$, $I(\infty) = a + b$, the normalized current, i.e., the fraction of charges that escape recombination at the given electric field, is
\begin{equation}
i(E) = \frac{I(E)}{a + b}.
\end{equation}
The prompt scintillation normalized to the value at $E = 0$, namely, 
\begin{equation}
Y(0) = a + x(a + b),
\end{equation}
results in the normalized prompt scintillation as
\begin{equation} \label{eq:lightyieldmodel1}
y(E) = 1 - \frac{i(E)}{1 + x(1+b/a)} = 1 - f_s i(E),
\end{equation}
where $f_s$ is the fraction of the prompt scintillation light from ionization.

Sato~\emph{et al.}~\cite{Sato1974} calculated the ratio of the number of direct excitations to ionizations in helium to be $0.45$ for electron recoils.  Among the excited atoms, $83\%$ are in the spin-singlet state and the remaining $17\%$ are in the spin-triplet states\cite{Sato1974}.  For the excimers that form on recombination of the ionization products, experiments indicate that approximately 50\% are in excited spin-singlet states and 50\% are in spin-triplet states~\cite{AdamsThesis}.  With $x = 0.37$ and $b/a = 1.0$, Eq.~\ref{eq:lightyieldmodel1} reduces to
\begin{equation} \label{eq:lightyieldmodel2}
y(E) \simeq 1 - 0.57  i(E).
\end{equation}
Thus, the zero-field normalized electron scintillation yield has a very simple dependence on the ionization current, and measurements of the latter have been made by Seidel \textit{et al.}~\cite{Seidel2014}.  However, there is some uncertainty in the prefactor, $f_s = 0.57$, in front of the ionization current term in Eq.~\ref{eq:lightyieldmodel2}.  This is primarily due to the uncertainty in the parameter $b/a$.  Conservatively, we can take the uncertainty of this parameter to be 50\%, and this would correspond to an estimated lower and upper bound on the prefactor of $f_s = 0.52$ and $f_s = 0.64$, respectively.  Hence, the sensitivity of Eq.~\ref{eq:lightyieldmodel2} to deviations of the parameter $b/a$ from 1.0 appears to be quite modest.  In the following, we will perform calculations using all three values of $f_s$ to gauge the robustness of the model to the imperfect knowledge of its parameters, and whenever not specified, the value of $f_s$ is taken to be 0.57.

\begin{figure}[h!]
	\includegraphics[width=\columnwidth]{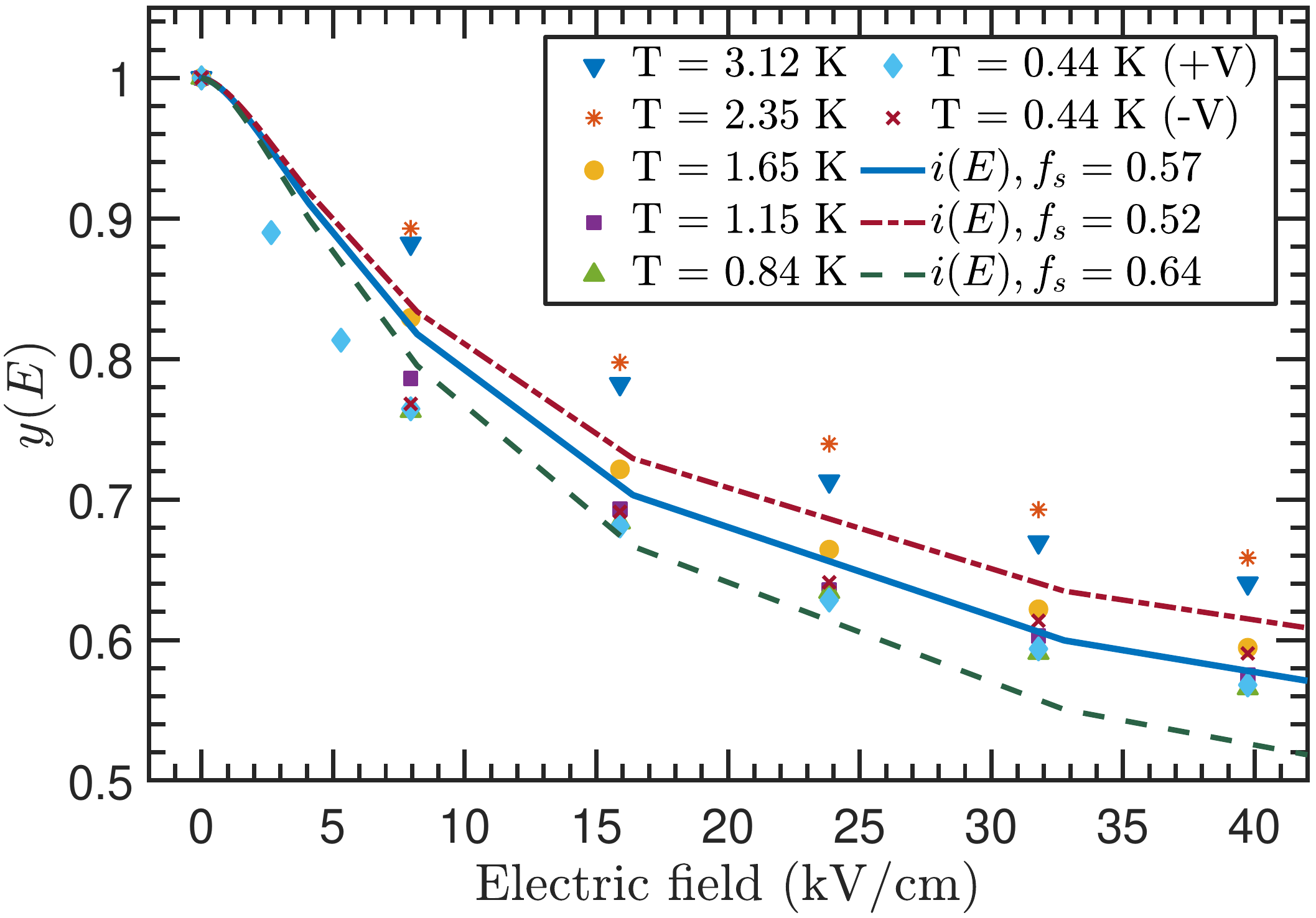}
	\caption{The normalized prompt light yield as a function of the applied electric field for 364~keV electrons from $^{113}$Sn.  The curves are the expected yield from Eq.~\ref{eq:lightyieldmodel2} with the ionization current from Seidel \textit{et al.}~\cite{Seidel2014} for different values of $f_s$. Refer to text for more details about the apparent temperature dependence.}
	\label{fig:lightyieldmodel}
\end{figure}

Utilizing $i(E)$ from Seidel \textit{et al.}~\cite{Seidel2014}, a comparison of the model prediction with data is shown in Fig.~\ref{fig:lightyieldmodel}.  There is fair agreement between the $f_s = 0.57$ curve and the low-temperature data ($<2$~K), but there are two immediate observations of note.  First, the model is not able to account for the observed temperature dependence of the light yield regardless of the value of $f_s$.  Additionally, even though the model appears to have better agreement with our low-temperature data, there still exists a discrepancy, particularly in the low field region (below $\approx 8$~kV/cm).  We address these observations in the following discussion.

\subsubsection{\label{sec:zerofieldtempdepend} Zero-field temperature dependence}

It is clear that the model is unable to account for the observed temperature dependence shown in Fig.~\ref{fig:lightyieldmodel}, and it is possible that this is merely a consequence of an inadequate model.  But from a different perspective, the reasonable agreement between model and data even with the use of an independently obtained ionization current measurement and numerical values of model parameters indicates there is at least some merit to this model.  Let us then suppose that there is an alternative explanation for the discrepancy and that our model has some veracity.  If such is the case, then a clue to the origin of the temperature dependence appears to lie in the zero-field scintillation data given that the normalization is performed with respect to them.  That an effect emergent at zero field could also appear in the finite field data is not surprising. The absolute scintillation yield shown in Fig.~\ref{fig:abslightyield}(b) is suggestive of this very possibility.  It shows that the absolute yield for different temperature datasets converge to approximately the same value with increasing field strength.  This is indicative of an effect that is manifested at zero or low fields but become greatly diminished, or possibly vanishes entirely, at higher fields.

If we consider the prompt signal at zero field as a function of temperature as shown in Fig.~\ref{fig:zerofieldPE}(b), then it is immediately apparent that the amount of detected light is much higher for the data measured at temperatures below the $\lambda$ transition.  For instance, the detected yield at 3.12~K is approximately 12\% lower than that measured at 0.44~K.  Perhaps, the behavior is merely a result of a liquid density temperature dependence?  However, a more careful examination reveals that a change in the liquid density cannot be the primary reason for the observed behavior.  Since our measurements are acquired with the liquid under a pressure of $\approx 600$~Torr, the difference in density between 0.44~K and 3.12~K is only about 3\%, and there are very small density differences for temperatures in the range of 0.44~K to 2.35~K (Table~\ref{tab:datasets}). Therefore, density effects alone cannot explain the observed temperature dependence of the zero-field data, particularly the $<3$~K data.

\begin{figure}[h!]
	\includegraphics[width=\columnwidth]{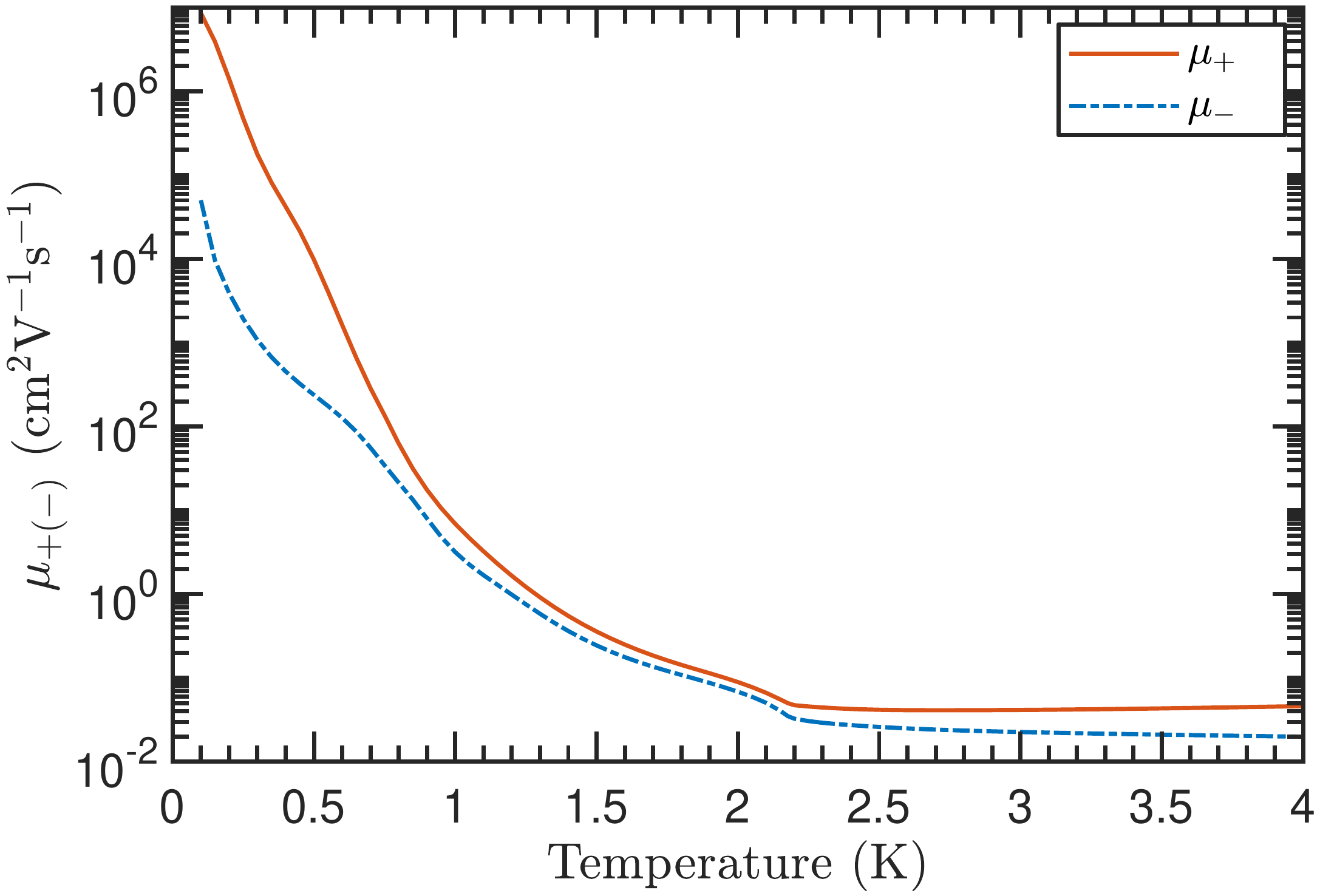}
	\caption{The zero-field mobility of positive and negative ions in LHe as a function of temperature from Donnelly and Barenghi~\cite{Donnelly1998}.}
	\label{fig:ionmobilities}
\end{figure}

\subsubsection{\label{sec:recomtime} Ion mobility and recombination time}

Of the many properties of LHe that are known to change with temperature, the mobilities of ions are of particular interest because they affect the recombination process.  As shown in Fig.~\ref{fig:ionmobilities} the zero-field mobility of both the positive and negative ions changes rather rapidly below the $\lambda$ transition \cite{Donnelly1998}.  The change here mirrors what is observed in the absolute scintillation yield of Fig.~\ref{fig:zerofieldPE}(b).  It is therefore conceivable that an effect arising from a temperature-dependent ion mobility may explain the zero-field temperature dependence of the electron prompt scintillation yield.

The mechanism by which the temperature-dependent mobility would affect the detected prompt light yield has to do with the finite recombination time for the thermalized ion pairs.  In this experiment, the integration time for the prompt pulse is set to 100~ns, so that scintillation light emitted by excimers formed through recombination after this time will not be accumulated in the prompt signal.  Qualitatively, the trend of the ion mobilities shown in Fig.~\ref{fig:ionmobilities} is not inconsistent with the observed lower light yield detected at higher temperatures; the mobilities are much smaller at these temperatures, so that the ions will take longer to recombine as compared to when the temperature is below the $\lambda$ transition, resulting in a reduced detected light yield for a given finite signal integration time window.

To illustrate this point more quantitatively, consider the simple case of a single pair of ions separated by distance $r_{0}$.  Once they have thermalized, the pair will drift towards one another due to their mutual Coulomb attraction.  Under the assumption of a field-independent mobility and ballistic ion motion, an estimate of the recombination time, $\tau_r$, is given by 
\begin{equation} \label{eq:estrecombtime3}
\tau_{r} = \frac{4\pi \epsilon_{0}\epsilon_{r}}{3q\mu}r_{0}^3 = \frac{4\pi \epsilon}{3q(\mu_{+} + \mu_{-})} r_{0}^3  ,
\end{equation}
where $\mu = \mu_{+} + \mu_{-}$ is the combined mobility, $q$ the electric charge, and $\epsilon = \epsilon_{0}\epsilon_{r}$ is the permittivity of LHe.   Williams~\cite{Williams1964} obtained the same estimate of the recombination time and also applied the Nernst-Einstein relation, $D = \mu k_{B} T /q $, to relate the mobility to the diffusion coefficient.

From Seidel~\textit{et al.}~\cite{Seidel2014}, the typical separation for ion pairs produced by an electron is $\approx 40$~nm with 10\% of the ion pairs having an initial separation greater than 100~nm.  Considering that the difference in the zero-field light yield between 0.44~K and 3.12~K is $\approx 12$\%, we take the latter separation distance as representative of the relevant length scale to consider.  The estimated recombination time at a separation of $r_0 = 100$~nm is $\tau_r \approx 38$~ns for $T = 3.12$~K but is $<1$~ns for temperatures below 1~K.  Hence, the estimate for $T = 3.12$~K is of the same order as the signal integration time.  Certainly, the assumption of ballistic motion is not entirely accurate because diffusion is present.  But in the more accurate description that includes the effects of diffusion, the true recombination time is longer than the estimate obtained from Eq.~\ref{eq:estrecombtime3}.  In fact, Ludwig~\cite{Ludwig1969} has shown that Eq.~\ref{eq:estrecombtime3} overestimates the recombination rate (i.e., underestimates the recombination time).  Thus, this estimate can be thought of as representing somewhat of a lower bound on the recombination time.

An upper bound on recombination time can be obtained by considering the case of diffusion dominated motion.  In such a case, the average time for a displacement, $r_0$, is
\begin{equation} \label{eq:estrecombtimediff}
\tau_{r, D} = \frac{r_{0}^2}{2(D_{+} + D_{-})},
\end{equation}
with $D_{+}$ and $D_{-}$ being the coefficients of diffusion for the positive and negative ions, respectively \cite{Jaffe1940}.  For $r_0 = 100$~nm and $T = 3.12$~K, the diffusion dominated recombination time is $\tau_{r,D} \approx 2900$~ns.  The true recombination time as well as the experimental integration time both lie within the calculated bounds, so the significance of this effect on the detected light signal cannot be immediately dismissed.

There are, of course, other effects that must also be considered in this analysis.  These include the field dependence of the mobility and the effect of vortex rings on ion motion.  The latter effect, however, appears not to be significant for temperatures close to $T_{\lambda}$; the positive ion does not become trapped in a vortex ring unless $T \lesssim 0.65$~K, and a negative ion will bind only up to $T \approx 1.7$~K \cite{Gamota1970}.  At a temperature of $T = 1.12$~K, the transit time for a negative ion to transverse some given distance is about a factor of four higher when it is trapped versus when it is free \cite{Gamota1970}.  But at this temperature, the estimated recombination time is $<1$~ns for $r_0 = 100$~nm, suggesting that the effect of vortex rings is mostly negligible to the present discussion.

A more rigorous treatment of the time-dependent problem of geminate recombination, that of isolated pairs of ions undergoing diffusive motion in their mutual Coulomb field, must start with the Debye-Smoluchowski equation,
\begin{equation} \label{eq:debyesmoluchowski}
\frac{\partial \rho (\vec{r}, t)}{\partial t} = \nabla D  \cdot \left[ \nabla \rho + \frac{\rho}{k_{B}T}\nabla U(r) \right],
\end{equation}
where $\rho(\vec{r},t)$ is the probability density of one ion relative to its partner at time $t$, $D = D_{+} + D_{-}$ is the diffusion coefficient, $T$ is the absolute temperature, and $U$ is the interaction energy for an isolated pair of ions~\cite{Rice1985}.

The well-known Onsager theory of geminate recombination~\cite{Onsager1938} corresponds to the solution of this equation in the limit of $t\rightarrow \infty$.  As such, the solution is only applicable in the steady-state situation and does not include any time dependence of the recombination process.  Nonetheless, approximate time-dependent solutions have been obtained by several authors \cite{Montroll1945, Mozumder1968, Abell1972, Rice1979, Pedersen1981, Parlange1981, Flannery1982, Green1984, Green1989}, and the full analytic solution was obtained by Hong and Noolandi for both the special and general cases of without and with an externally applied electric field \cite{HongNoolandi68,HongNoolandi69}.  Unfortunately, their full analytic solutions are found in Laplace transform space and as a consequence are immensely complicated and difficult to evaluate numerically, so that application to the analysis of experimental data is not entirely straightforward.  In the discussion that follows, an approximate solution in the time domain developed by Green~\textit{et al.}~\cite{Green1989}, which is applicable to the situation in which the initial ion pair separation distance is significantly smaller than the Onsager radius and the medium has a low permittivity, is used for the analysis of our data.

\subsubsection{\label{sec:survivalprob} Recombination survival probability}

\begin{figure}[!h]
	\includegraphics[width=\columnwidth]{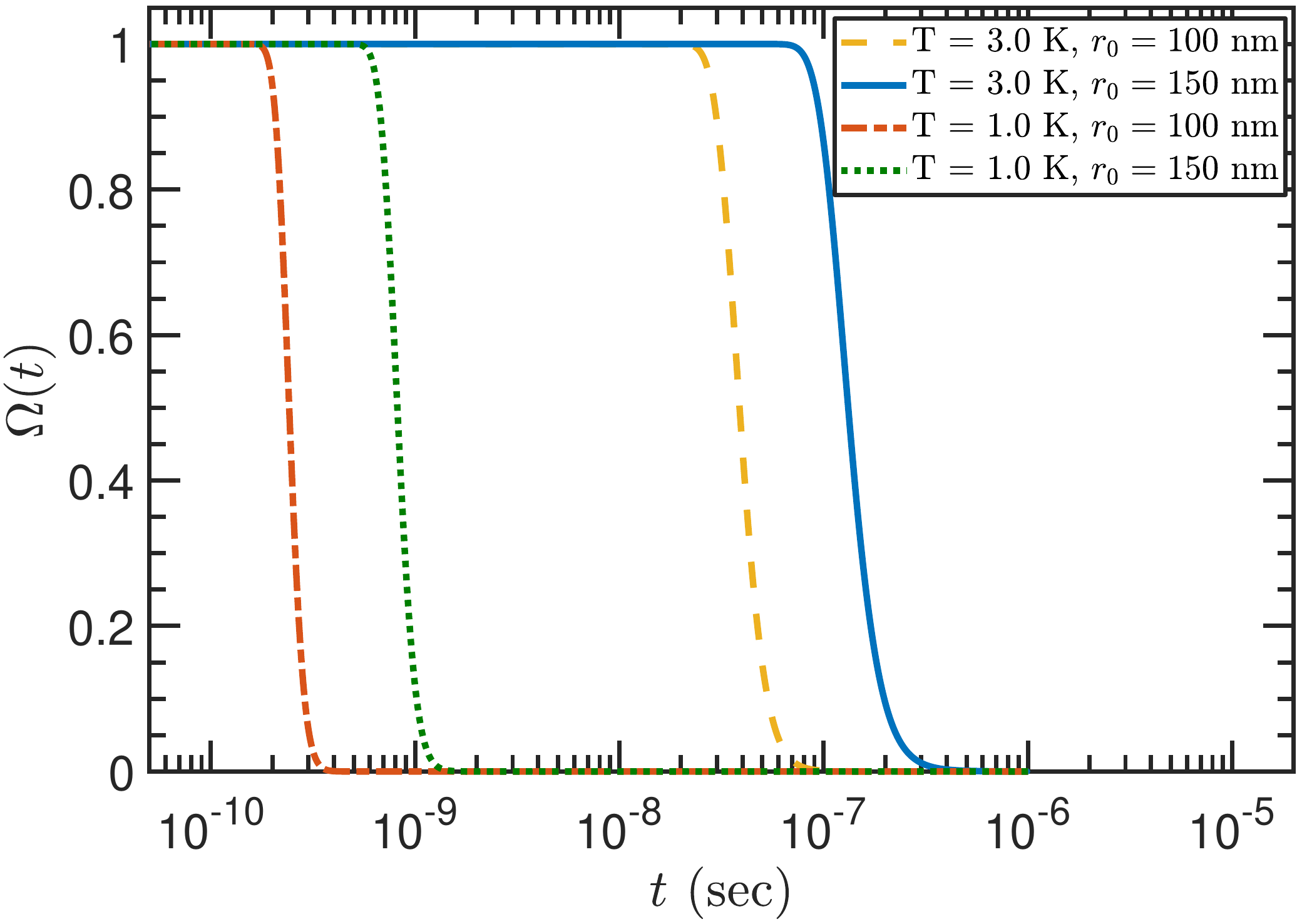}
	\caption{The recombination survival probability as a function of time for two different values of $r_0$ at $T = 1.0$~K and  $T = 3.0$~K.  The curves are calculated using the approximate solution developed by Green~\textit{et al.}\cite{Green1989}.}
	\label{fig:approxsolutions}
\end{figure}

When discussing recombination, the quantity of particular interest is the survival probability, $\Omega(r,t,T)$, which is the fraction of particles that have not recombined in the system at time $t$, temperature $T$, and initial separation $r$.  Shown in Fig.~\ref{fig:approxsolutions} is the calculated survival probability using the Green~\textit{et al.} approximate solution \cite{Green1989} to the Debye-Smoluchowski equation.  The probability is calculated for two different initial ion separations of 100~nm and 150~nm at a temperature of 1.0~K and 3.0~K by making use of the mobility data from Donnelly and Barenghi~\cite{Donnelly1998}.  The results indicate that the survival probability at 3.0~K approaches 1.0 for an initial separation of $r_{0} = 150$~nm with $t = 100$~ns, the signal integration time in the experiment.  At 1.0~K, the survival probability is zero for both initial separations when $t = 100$~ns.  Notably, the survival probability changes rather rapidly, falling from one to zero within a single decade of time.  This behavior is implied by the power-law dependence of the initial separation in the recombination time from our earlier estimate from Eq.~\ref{eq:estrecombtime3}. 

The implication of this is that any geminate pairs with an initial separation greater than $\approx 150$~nm will not contribute to the prompt scintillation signal for the chosen integration time at 3.0~K but will contribute at 1.0~K.  The degree to which the recombination of geminate pairs contributes to the signal will depend on the distribution of their initial separation.

\subsubsection{\label{sec:chargedist}Thermalization distribution}

A determination of the temperature dependence of the light yield requires knowledge of the thermalization distribution.  The initial separation of an ion pair is also typically referred to as the thermalization length, so we will adopt this latter terminology in the discussion that follows.  The secondary electrons produced in the wake of an ionization track will have a distribution of thermalization lengths relative to their geminate partners, $N(r)$, and in general the thermalization length distribution is also a function of temperature.  The survival probability averaged over this distribution is given by 
\begin{equation} \label{eq:probchargedistr}
\overbar{\Omega}(t,T) = \int_{0}^{\infty} \Omega(r,t,T) N(r) 4 \pi r^{2}dr.
\end{equation}
It then follows that the normalized detected light yield as a function of temperature is
\begin{equation} \label{eq:lightyieldprob}
y(t,T, E = 0) = f_{r} \left [ 1 - \overbar{\Omega}(t,T) \right ] + f_{ex} ,
\end{equation}
where $f_{r}$ and $f_{ex}$ are the fraction of scintillation due to recombination and excitation at zero field, respectively.  Implicit in Eq.~\ref{eq:lightyieldprob} is the assumption that any time dependence in the excitation component of the signal is negligible compared to the experimental integration time.  This is supported by experimental observation that the singlet excimers radiatively decay within 10~ns of the initial ionization event \cite{McKinsey2003}, a significantly shorter timescale than the signal integration time of our experiment.

As discussed below, $N(r)$ can be determined from the ionization current data, or equivalently the scintillation yield dependence on the  electric field.  We will determine $N(r)$ from our scintillation data.  However, fitting the light yield for the two highest-temperature datasets to derive the distribution is not possible because the presence of the recombination effect on the zero-field signal directly affects the field dependence.  For this reason, we will use the 0.44~K data to derive the distribution.

The thermalization distribution is determined by using the analytic method developed by Seidel~\textit{et al.}~\cite{Seidel2014}.  The ionization current, $i(E)$, is determined from the normalized detected light yield, $y(E)$, from Eq.~\ref{eq:lightyieldmodel2}.  Then, $N(r)$ is obtained from the relation
\begin{equation} \label{eq:currentchargedistrequation}
N(r) = \frac{4 \pi^{1/2} \epsilon_{0}^{3/2} E^{5/2}}{e^{3/2}} \frac{d}{dE} \left( i + E \frac{di}{dE}   \right)
\end{equation}
which applies to motion in a viscous medium~\cite{Seidel2014}.  
 
For fitting the detected light yield, the use of a spline polynomial model is not preferred because it would restrict the fit to the range of the data.  Furthermore, the sparseness of our data points makes such a fit model unreliable.  Instead, a more reliable fit can be obtained by specifying a functional form.  For this, we choose the simple analytic expression proposed by Boag~\cite{Boag1950} and rediscovered by Thomas and Imel~\cite{Box1987} in their efforts to determine the fraction of charges escaping initial recombination.  

We note that there are possible objections to using this model given its lack of physical justification.  However, its experimental success and general applicability to both columnar and cluster recombination show its usefulness, even if only as an empirical model.  But it is also important to note that although the model's use provides a straightforward and convenient means for obtaining the sought after thermalization distribution, our use is limited to only its \emph{functional form} and does not represent a means to obtain a determination of any physical parameters.

\begin{figure}[htb!]
	\includegraphics[width=\columnwidth]{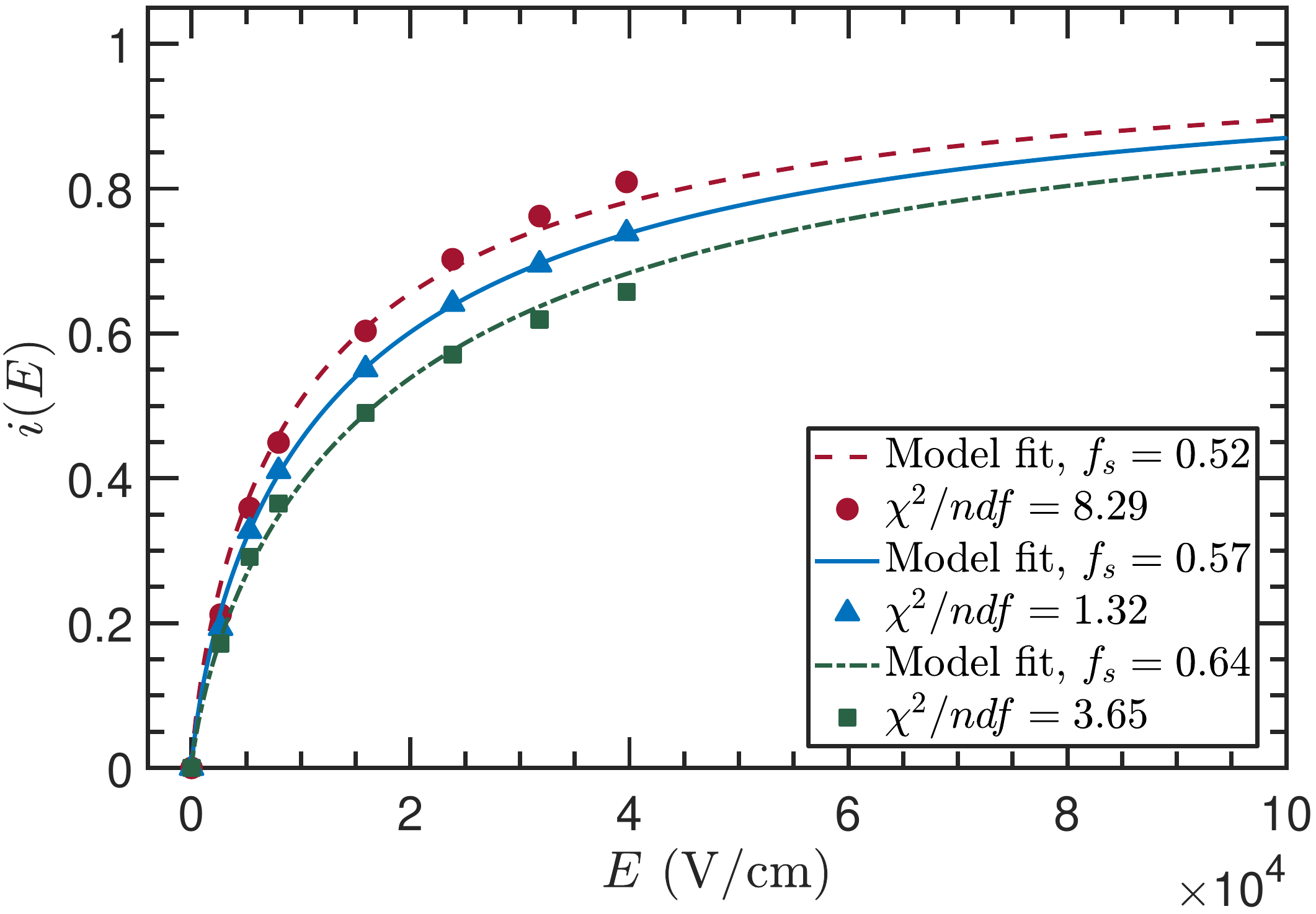}
	\caption{Fits of the normalized current derived from the scintillation measurements and the model in Eq.~\ref{eq:lightyieldmodel2} for the $T = 0.44$~K data with a function of the form proposed by Refs.~\cite{Boag1950} and \cite{Box1987}.}
	\label{fig:funcfitnormyield}
\end{figure}

The fit of the normalized current for the $T = 0.44$~K dataset using the chosen functional form is shown in Fig.~\ref{fig:funcfitnormyield}.  The goodness of fit is indicated by the reduced-$\chi^2$ value ($\chi^2$/ndf).  We note that the data points used in these fits are derived from the midpoints spanned by the positive and negative polarity datasets.  For the two field data points ($\approx$~2.5 and 5~kV/cm) in the positive polarity set that are not present in the negative polarity set, the values are simply taken from the positive polarity set without alterations.  The polarity effect at those two points is negligible as indicated by Fig.~\ref{fig:polarityeffect}.  Fitting of the ionization current is done for the three values of $f_s$, and the best fit is obtained for $f_s = 0.57$.

\begin{figure}[htb!]
	\includegraphics[width=\columnwidth]{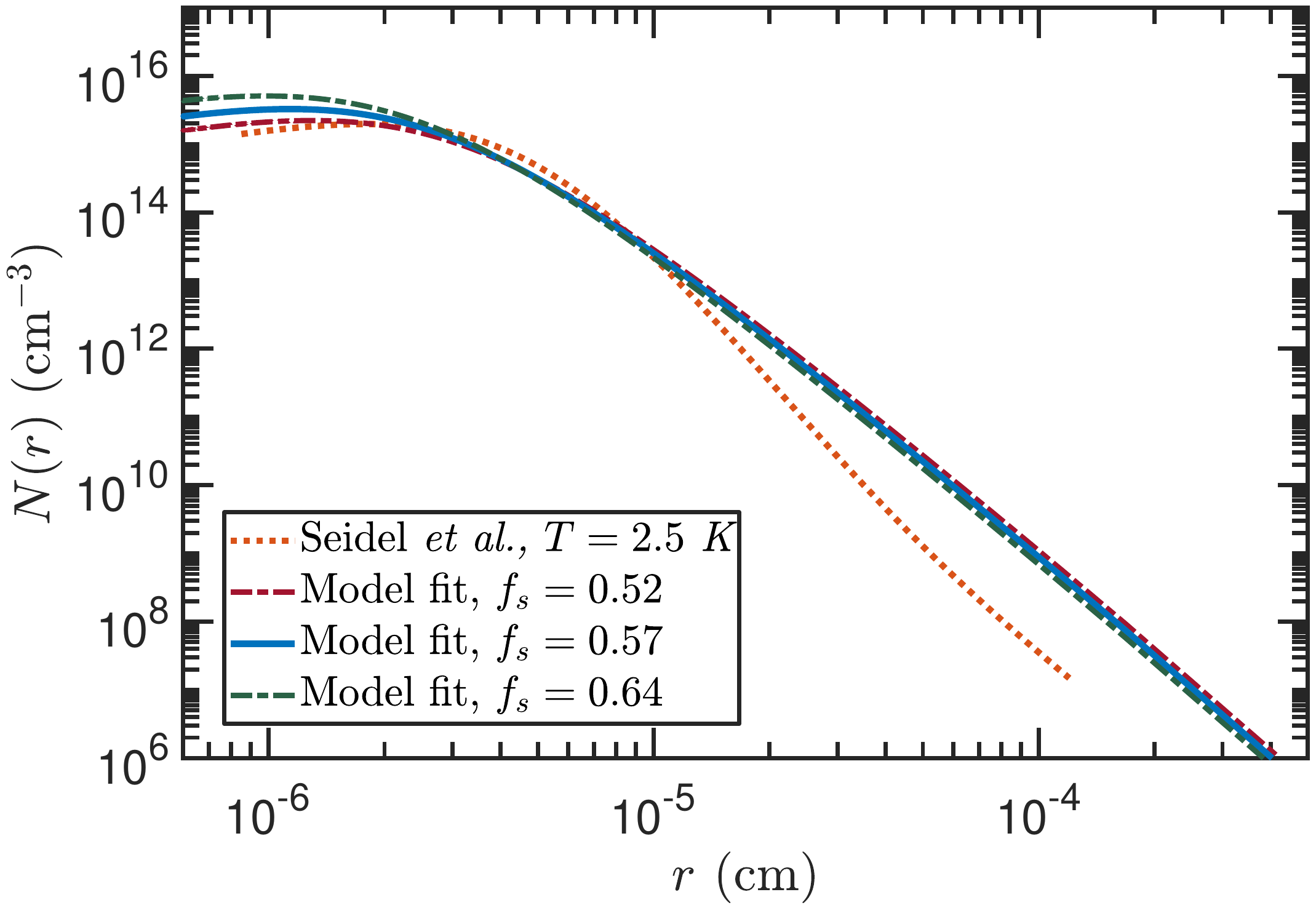}
	\caption{The thermalization distribution derived from scintillation measurements in this work for different values of $f_s$ and the distribution from Seidel \textit{et al.}~\cite{Seidel2014}.}
	\label{fig:chargedistr}
\end{figure}

In Fig.~\ref{fig:chargedistr}, the distributions obtained from the fit of the $T = 0.44$~K dataset with different $f_s$ values along with the distribution derived from the ionization current measurements of Seidel~\textit{et al.}~\cite{Seidel2014} are shown.  The derived distributions are similar regardless of the value of $f_s$, and so the shape of the distribution appears to be robust against uncertainty in the parameter $b/a$ of the model.  The important feature to highlight is the presence of a much heavier tail in the distributions derived from the scintillation measurements in this work.  In particular, the fraction of charges with separation greater than $100(150)$~nm is 25\%(15\%) for the distribution ($f_s = 0.57$) derived in this work compared to only 10\%(3\%) for the distribution from Seidel~\textit{et al.}~\cite{Seidel2014}.  At separations greater than 100~nm, our distribution has approximately an $r^{-4}$ dependence.  The probability density function for a thermalization length $r$ is $N(r)r^2$ and thus has a $r^{-2}$ dependence. A distribution that has a tail with such a dependence is often referred to as a ``fat" or ``long" tail distribution.  The random motion of an electron as it thermalizes in the liquid through elastic scattering with the surrounding helium atoms will produce such a distribution if the distance an electron travels between scatters is highly variable~\cite{Seidel2014}.  That situation is described by a process called a Levy walk, where the distribution of step sizes is $f(x) \sim |x|^{-(1+\alpha)}$ with $0 < \alpha < 2$ \cite{Ibe2013}.

The distribution derived from our scintillation measurements suggests there is a larger fraction of ion pairs separated by distances greater than 100~nm than what is obtained from the distribution of Seidel~\textit{et al.}~\cite{Seidel2014}.  In fact, this result is expected if the field dependence of the electron scintillation is well described by the model proposed in Sec.~\ref{sec:model}.  The reason for this expectation is that the curves from Fig.~\ref{fig:lightyieldmodel} are suggestive of a heavier tailed thermalization distribution; the data show a larger yield reduction at low fields than what the model predicts with the ionization current from Seidel~\textit{et al.}~\cite{Seidel2014}.  However, it is important to note that this does not necessarily suggest that the distribution from their work is in error.  Rather, it is likely that their distribution is not applicable to the scintillation data from this work due to differences in experimental setups.  In particular, the energies of the electron sources used in the two experiments are quite different (mean energy of 17~keV for $^{63}$Ni vs 364~keV for $^{113}$Sn ), and so the distribution of separation distances of the thermalized charges need not necessarily be the same because the initial energy and spatial distribution of the secondary electrons are not necessarily identical.  We will expand on this point in more detail in Sec.~\ref{sec:distenergydepend}.

\begin{figure}[htb!]
	\includegraphics[width=\columnwidth]{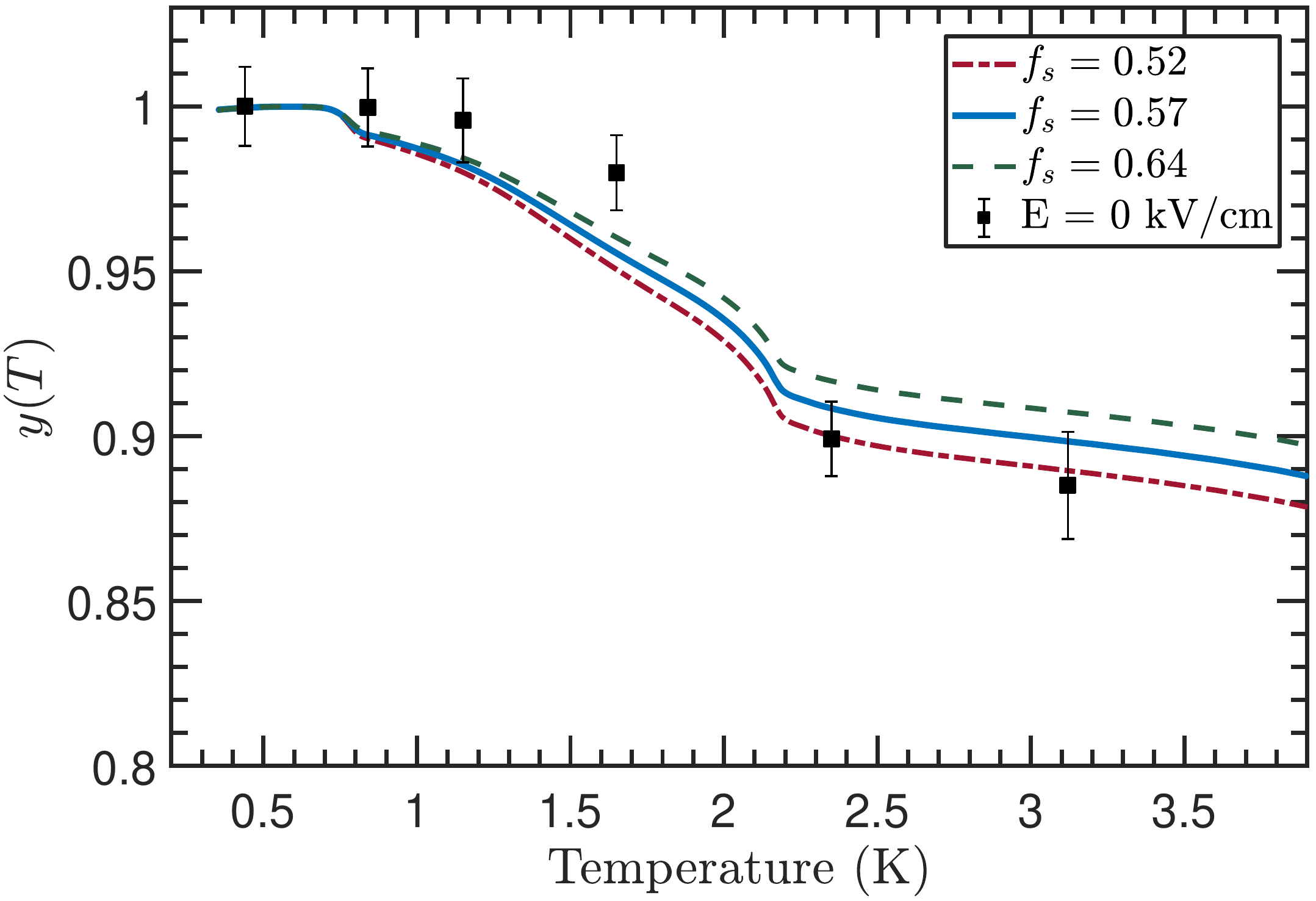}
	\caption{The normalized scintillation yield as a function of temperature normalized to the yield at 0.44~K (black squares), and the yield obtained from calculations of the recombination probability for the thermalization distributions derived from model fits with different values of $f_s$.}
	\label{fig:yieldapproxsol}
\end{figure}

Utilizing the model derived thermalization distributions, the predicted light yield as a function of temperature at zero field as given by Eq.~\ref{eq:lightyieldprob} is shown in Fig.~\ref{fig:yieldapproxsol}.  The distributions are obtained from a fit of the scintillation data at 0.44~K.  The fraction of prompt scintillation light due to recombination and excitation is taken to be $f_s$ and $1-f_s$, respectively.  Corrections for density effects on the thermalization distribution are included in the calculations.  This is done by letting $r\rightarrow r(\rho_0/\rho_T) $, where $\rho_0 $ is the density at 0.44~K and $\rho_{T}$ is the density at temperature $T$.  Additionally, the data points incorporate changes to the photon geometric acceptance onto the TPB wavelength shifting coating due to density/temperature effects, and these are accounted for with simulations.  

The calculated temperature dependence of the zero-field yield reproduces the general trend of the data.  Uncertainties in the value of the prefactor $f_s$ does not appear to significantly alter the behavior of the yield with temperature.  The lack of complete agreement between calculations and data may be due to the presence of other effects besides finite recombination time.  Perhaps, another temperature-dependent effect influencing the scintillation yield exists, one which is not accounted for by our model.  Moreover, it must also be mentioned that in the discussion thus far we have assumed a field-independent mobility.  However, experiments show that this assumption is only valid for fields up to a few 100~V/cm, beyond which point the mobility decreases.  When the ion-pair separation is less than $\approx 100$~nm, the field strength is $>1000$~V/cm and the drift velocity of the ions at low temperatures may approach the Landau limit ($\approx 60$~m/s), above which the mobility reduces with increasing field strength \cite{MayerReif1958}.  But when the drift velocity is at the Landau limit, the time to transverse the final 100~nm is only a few ns, so this effect appears negligible when compared to the signal integration time.  At higher temperatures, a decrease in mobility at high fields may still play a significant role.

To summarize: In Sec.~\ref{sec:model}, we introduced a model that related the scintillation yield to the ionization current.  We then compared the observed effect of an electric field on the scintillation yield measured in this work to the model prediction that used the ionization current data from Ref.~\cite{Seidel2014}. We found that our data for the normalized scintillation yield for $T < 2$ K has fair agreement with the model, but the apparent temperature dependence observed is not immediately explained.  In Sec.~\ref{sec:zerofieldtempdepend}, we pointed out that the apparent temperature dependence of the normalized scintillation yield as function of field can, however, be attributed to the temperature dependence of the zero-field scintillation yield, which implied that the model is not necessarily inadequate. Section~\ref{sec:recomtime} discussed the possible effect that the temperature-dependent mobility of positive and negative ions has on the detected zero-field scintillation yield when the signal integration time is finite. In Sec.~\ref{sec:survivalprob}, we presented a more refined treatment of the effect using an approximate solution of the Debye-Smoluchowski equation. Before applying this solution, we determined the ionization current and ion thermalization distribution from our low-temperature data (0.44 K) in Sec.~\ref{sec:chargedist}.  Using this information together with the approximate solution, we then calculated the effect of the temperature-dependent mobility on the zero-field scintillation yield for the signal integration time of 100 ns in this experiment. The calculated effect is shown to agree well with our observed zero-field temperature dependence.  Thus, there is consistency between the model and the features observed in our measurements after accounting for the effect of finite recombination time on the detected scintillation signal.

The preceding analysis shows that the ionization current and thermalization distribution determined from our scintillation data exhibit clear differences to those from Ref.~\cite{Seidel2014}. We discuss the possible reason behind this disparity and its implications in Sec.~\ref{sec:distenergydepend}.

\subsection{\label{sec:distenergydepend} Energy dependence of thermalization distribution}

The spatial distribution of thermalized secondary electrons with respect to their geminate partners depends on the initial energy distribution of the electrons and the energy loss processes they undergo in the medium.  For instance, a 10~eV electron is estimated to require on the order of $10^4$ collisions before being thermalized in a sphere of approximately 100~nm from its positive ion partner \cite{Seidel2014}.  In principle, knowledge of the initial energy distribution and all the energy loss processes allows for the determination of the spatial distribution of the thermalized electrons. However, in practice, that determination is not so straightforward.  By restricting our attention to only the energy distribution of secondary electrons, the problem is made much more tractable while still providing useful information about the energy dependence of the thermalization distribution.

\begin{figure}[htb!]
	\includegraphics[width=\columnwidth]{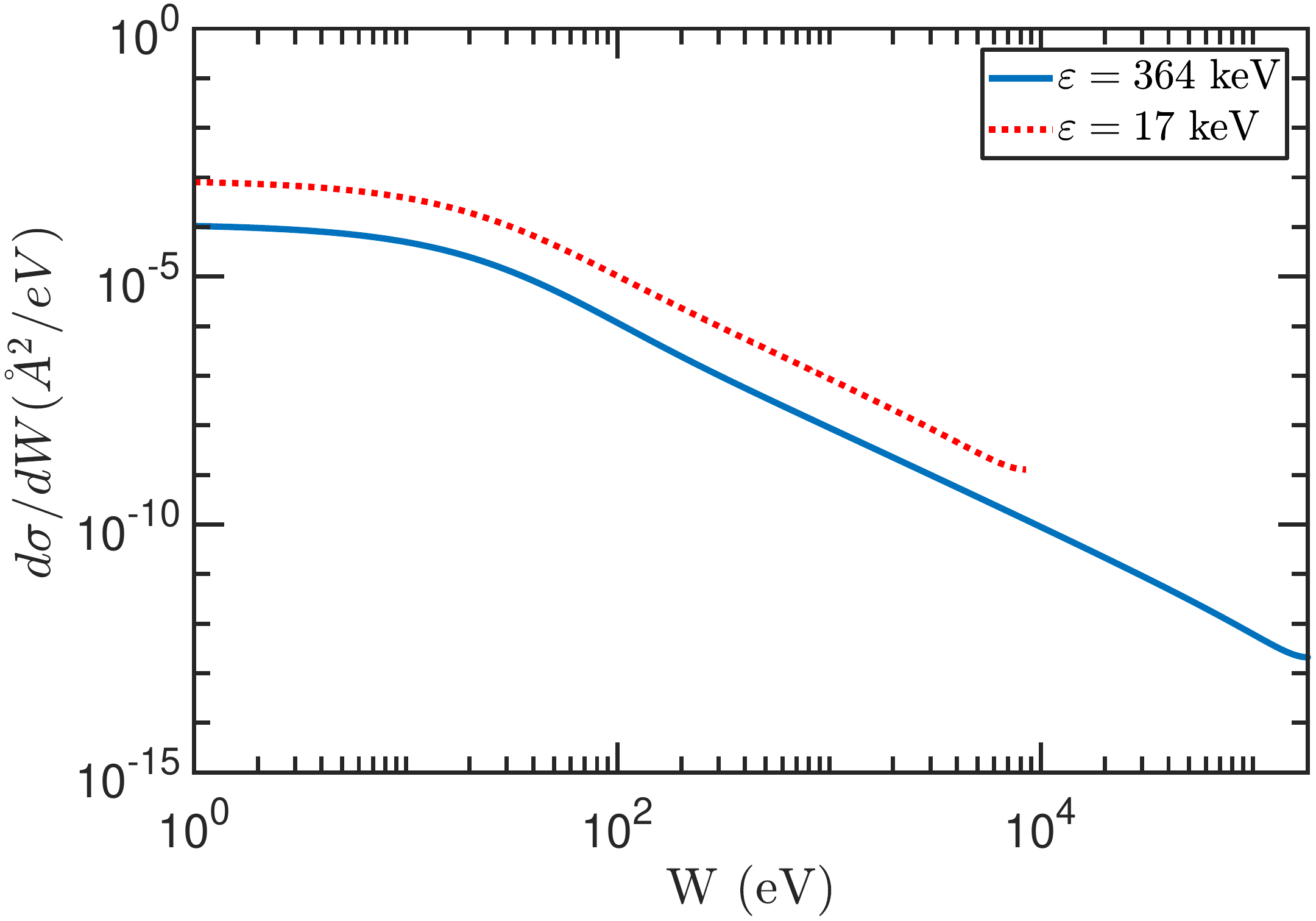}
	\caption{The single differential ionization cross-section for primary electron energies of $\varepsilon = 17$ and 364~keV in helium calculated using the relativistic extension of the binary-encounter-dipole model from Ref.~\cite{Kim2000}.}
	\label{fig:diffcrosssection}
\end{figure}

To start, let us consider the single differential ionization cross-section for electron impact, $d\sigma/dW$, as a function of the secondary electron energy, $W$, in helium. This is plotted for two different primary electron energies in Fig.~\ref{fig:diffcrosssection} . The formulas from Ref.\cite{Kim2000} used for this calculation are valid for relativistic primary electron energies and are an extension of the binary-encounter-dipole model originally proposed by Kim and Rudd~\cite{Kim1994}.  The lower energy curve represents the energy distribution of secondary electrons from the $^{63}$Ni source (mean primary electron energy of 17~keV) used in the ionization current measurements of Seidel \textit{et al.}~\cite{Seidel2014}, while the 364~keV curve represents the distribution of the $^{113}$Sn source used in the scintillation measurements from this work.

Interestingly, the shape of the energy distribution of secondary electrons is nearly identical for the two primary electron energies being considered as shown in Fig.~\ref{fig:diffcrosssection}.  The main difference is that the lower primary energy distribution is shifted upward toward higher cross-sections relative to the other distribution.  Therefore, this would seem to imply that the thermalization distribution is independent of the energy of the primary electron.  However, even though the shape of the energy distribution of secondaries is similar, the relative shift between the two is key to understanding the energy dependence of the thermalization length distribution which is ultimately related to the difference in the ionization density and the mean separation distance between ion-pairs.

Before discussing the importance of the ionization density, let us further inspect the shape of the energy distribution by considering the fraction of the secondary electrons having energies in the range of 10 to 19.8~eV.  The estimated thermalization length for an electron with an energy of 10~eV is 100~nm \cite{Seidel2014}, and 19.8~eV corresponds to the energy of the first excitation level in helium, below which electrons can only lose energy through the inefficient process of scattering with helium atoms.  To estimate the fraction of secondary electrons that have a thermalization length greater than 100~nm, we integrate the distribution in Fig.~\ref{fig:diffcrosssection} from $W = 10$~eV to $W = 19.8$~eV and take the ratio of the result to the total ionization cross-section.  This results in $\approx 21$\% and $20$\% of secondary electrons being thermalized at separations $>$ 100~nm for a primary electron energy of 364 and 17~keV, respectively.  These values represent only a lower bound because very high energy secondary electrons can further ionize, producing tertiary electrons, some of which will have energies in the range of 10 to 19.8~eV.  Interestingly, the fraction obtained from this estimate is similar to the value (25\%) obtained from the distribution derived from the model fit of our scintillation data, but at the same time is significantly higher than the value (10\%) obtained from the distribution by Seidel~\textit{et al.}~\cite{Seidel2014}.

The difference between the thermalization distribution derived from the scintillation data in this work and that obtained from the ionization current measurements of Seidel~\textit{et al.}~\cite{Seidel2014} is likely the result of the energy of the electron source used in the experiment.  The electrons from the $^{63}$Ni electron emitter used in Seidel~\textit{et al.}~\cite{Seidel2014} have an end point energy of 66~keV and a mean energy 17~keV.  Taking the value of 17~keV as the characteristic electron energy of the $^{63}$Ni source and a $W$ value of 43~eV in helium, the electron average range is $\approx 4.3 \times 10^{-3}$~mm in LHe and the average separation distance between adjacent ion pairs, $\bar{x}$, is $\approx 100$~nm.  By comparison, the 364~keV electrons from the $^{113}$Sn conversion source used in this work have an average range of $\approx 7$~mm, corresponding to an average separation between ion pairs of $\approx 840$~nm.  Considering that the thermalization length for an electron is a few 10s to 100s of nm, there is a nonnegligible chance for an electron, once thermalized, to become paired/matched with a new positive ion partner, one in which it did not originate from, when the average separation between ion-pairs is much smaller than the thermalization length.  This shifts the distribution $N(r)$ to shorter $r$ values. On the other hand, when the energy of the electron is such that the average separation between ion pairs is much greater than the average thermalization length, the exchange of ion partners does not occur.

\begin{figure}[htb!]
	\includegraphics[width=\columnwidth]{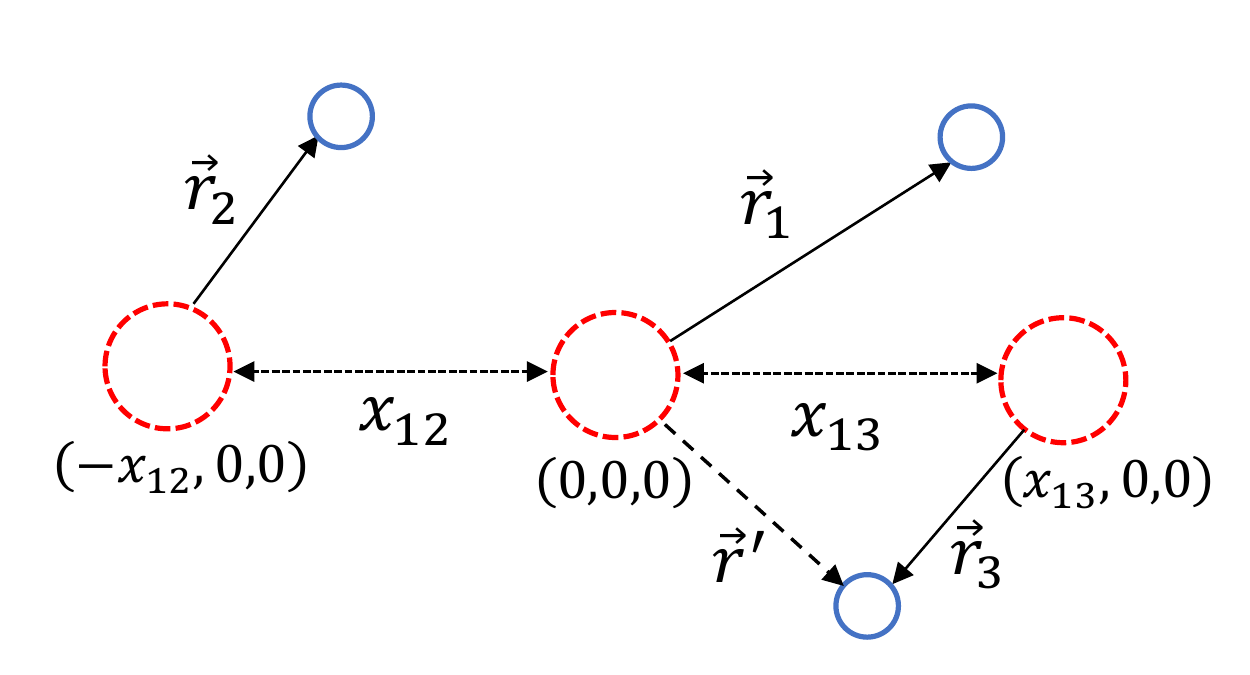}
	\caption{A diagram showing the model setup used in the simulation of the thermalization distribution and its dependence on the primary electron energy. Red-dashed circles represent the ion locations and blue circles the associated electrons.}
	\label{fig:simchargedistrdiagram}
\end{figure}

We further explore such an effect on the thermalization distribution through Monte Carlo simulations.  Consider three electron-ion pairs as shown in Fig.~\ref{fig:simchargedistrdiagram}.  The first pair is located at the origin while the second and third pairs are located at distances $x_{12}$ and $x_{13}$ from the first pair, respectively.  The pairs are arbitrarily chosen to lie on the $x$-axis, and the $x$'s, which are the separations between ion-pairs, are sampled from an exponential distribution.  The mean of this distribution, $\lambda$, is dependent on the energy of the primary electron.  The appropriate value for $\lambda$ can be set in the simulation for an arbitrary primary electron energy.  For the case of a 17~keV electron in LHe, $\lambda \approx 100$~nm.  At the location of each ion-pair, a random isotropic vector direction, $\hat{r}$, is chosen.  The length of the vector is sampled from the scintillation data derived distribution, $N(r)r^2$.  These two randomly generated quantities represent the direction and thermalization length of the electron.  The distance between the positive ion at the origin and the three thermalized electrons are then calculated, and the minimum distance, $r'$, is determined.  The process is repeated many times to accumulate a distribution of $r'$.

\begin{figure}[htb!]
	\includegraphics[width=\columnwidth]{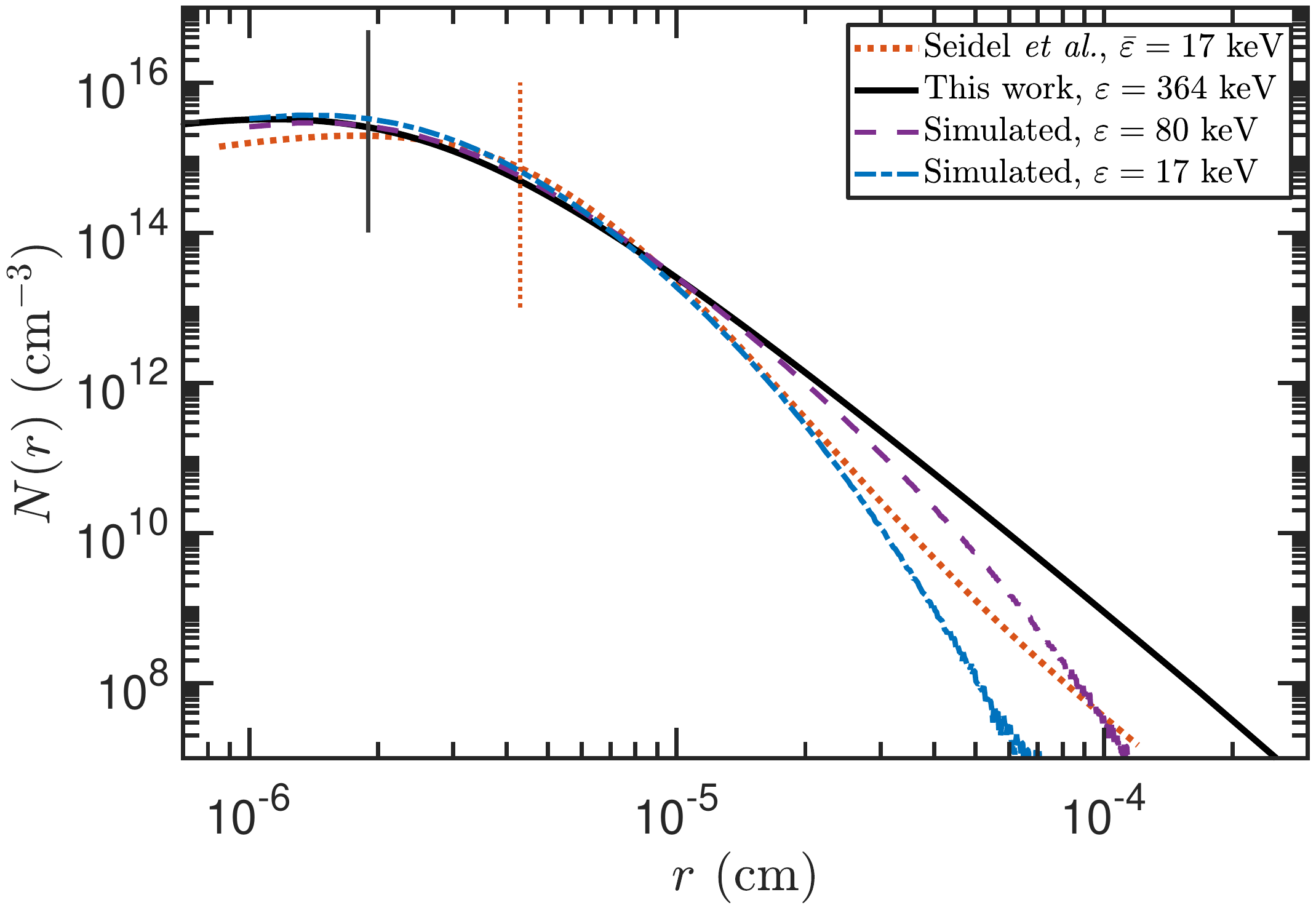}
	\caption{The simulated thermalization distributions for 17~keV (blue dash-dotted) and 80~keV (purple dashed) primary electrons and the distributions from Seidel \textit{et al.}~\cite{Seidel2014} (red-dotted ) and this work ($f_s = 0.57$) (black solid).  The red dotted and black solid vertical line segments at $r = 4.3\times 10^{-6}$~cm and $r= 1.9 \times 10^{-6}$~cm denote, respectively, the reach of the measurement of the corresponding experiment. Refer to text for further details. }
	\label{fig:simchargedistrs}
\end{figure}

The results of the simulations are shown in Fig.~\ref{fig:simchargedistrs} along with the thermalization distribution from Seidel~\textit{et al.}~\cite{Seidel2014} and the one derived from the scintillation measurements in this work with $f_s = 0.57$.  There is relatively good agreement between the simulated distribution and the one from Seidel~\textit{et al.}~\cite{Seidel2014} from $r \approx 2\times 10^{-5}$~cm to $r \approx 4.3\times 10^{-6}$~cm, with the latter separation corresponding to the highest field measurement in their experiment.  Here, $r$ is related to the applied field by $r = (e/4\pi \epsilon_0 E)^{1/2}$.  The small vertical separation between the two in this region is likely due to a normalization difference, whereas the wider separation at large thermalization distances is most probably the result of the simplistic nature of the simulations.  One such simplification is the use of a single primary electron energy, which does not describe the full emission spectrum of the $^{63}$Ni source which has an end point energy at 66~keV.  As a result, the high-electron-energy component is missing from the simulation, and this can be seen as the steeper tail at large thermalization distances in Fig.~\ref{fig:simchargedistrs}.  Further highlighting this point is the simulated distribution for 80-keV electrons, which shows an enhancement of the tail with primary electron energy.

We note that these rudimentary simulations are only meant to illustrate some qualitative features of the thermalization distribution and its energy dependence.  The complexity necessary to accurately simulate the distribution for an arbitrary electron energy is considerable and well beyond the scope of this work.  But in principle, once the thermalization distribution is known, it is a straightforward excercise to determine the scintillation and ionization current yields as a function of applied electric field.  

\begin{figure}[htb!]
	\includegraphics[width=\columnwidth]{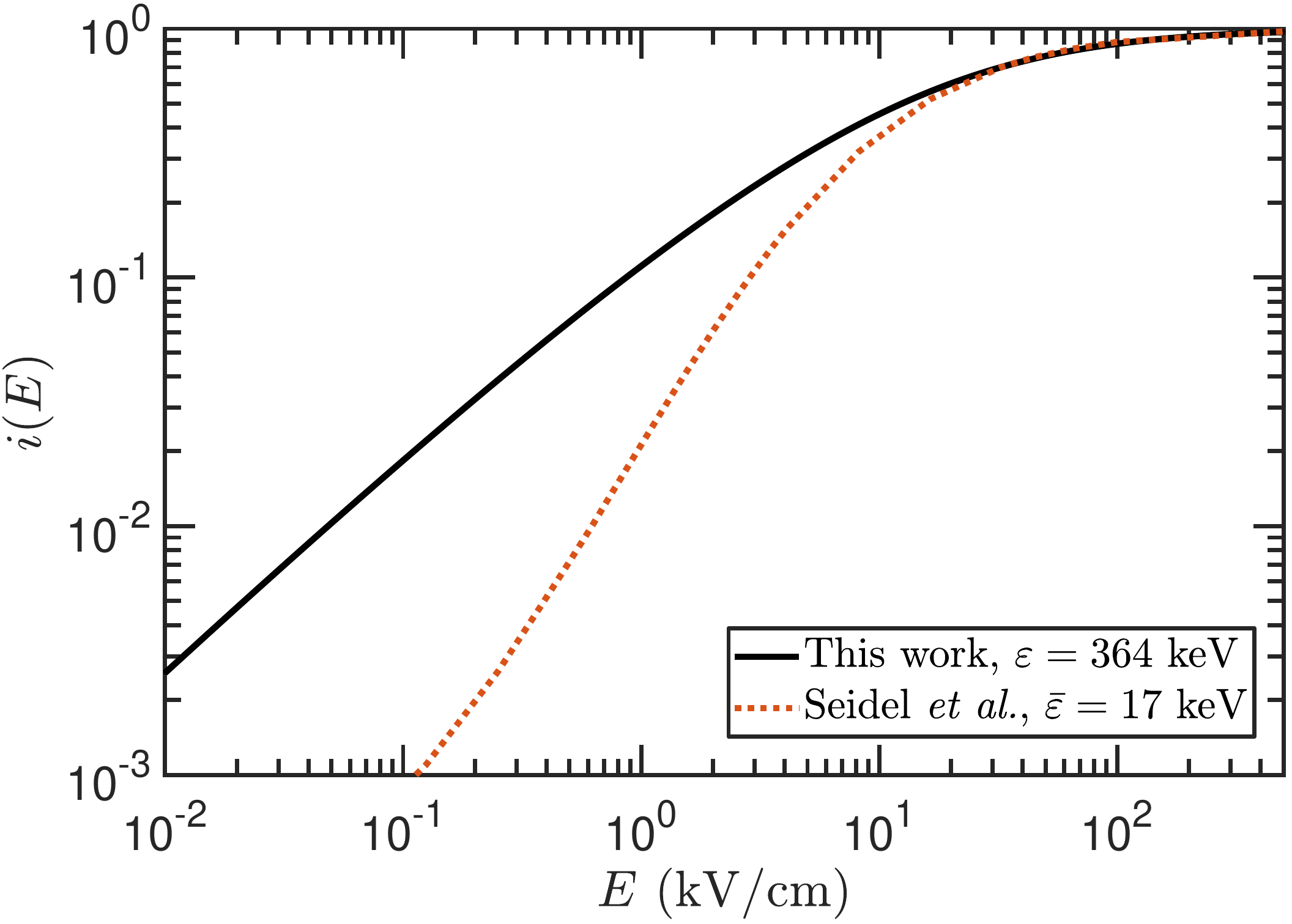}
	\caption{Comparison of the ionization current as a function of applied electric field as determined from the scintillation measurements in this work (solid black line) against the direct ionization measurements of Ref.~\cite{Seidel2014} (red-dotted line).}
	\label{fig:currentvsfield}
\end{figure}

Figure~\ref{fig:currentvsfield} shows a comparison of the ionization current yields deduced from our scintillation measurements for $f_s = 0.57$ to the direct ionization current measurements from Ref.~\cite{Seidel2014}.  Just as the thermalization distribution has an energy dependence, the same dependence is present in the ionization yield.  This has important implications for particle detection in that the measured ionization yield at one energy cannot be assumed to be applicable to the entire range of the measurements in the experiment.  For example, with an applied field of 1~kV/cm, approximately 10\% of charges are extracted from the ionization produced by a 364~keV electron, but this decreases to $\approx 2$\% when the mean energy is 17~keV.  At even lower energies, the yield should further decrease and approach that for heavier ionizing particles.  However, a determination of the precise level of reduction becomes untenable with the present analysis because the conditions for geminate recombination are no longer met when the charge density is sufficiently high, as is the case for electron energies of a few keV. 

The ionization yields are shown to converge when the electric field increases.  The characteristic field strength for this is approximately the strength required to pull apart a pair of ions separated by the mean length of the thermalization distribution, above which the contribution to the total ionization yield from the tail of the distribution becomes less significant.  This behavior is perhaps the reason for the improvement in particle discrimination observed for noble liquid detectors with increasing drift field in the detection volume~\cite{Aprile2006, Washimi2018}.

\subsection{Prompt pulse shape and particle identification}

The existence of a temperature-dependent recombination time in LHe may provide for a potential application in particle detection and identification.  As we have discussed, this dependence results from the thermalization distribution of the ion-pairs produced in the wake of an ionizing particle and the influence of a temperature-dependent mobility on the recombination of the ions.  As it is known that the ionization density produced is related to the properties of the charged particle, the effect is possibly very different for minimally ionizing particles (e.g., fast electrons) as compared to heavy ionizing particles (e.g., $\alpha$ particles and heavier nuclei).  For instance, the recombination time for $\alpha$ particles is estimated to occur on a timescale of only $\approx 0.1$~ns \cite{Ito2012} whereas that for electrons is temperature dependent and can be much longer.  The consequence of this is that the shape of the prompt scintillation signal may encode in it the identity of the type of ionizing particle.  Such an analysis of the scintillation signal shape for particle identification is commonly employed in detectors using solid-state or liquid scintillators~\cite{Knoll}.

\begin{figure*}[]
		\includegraphics[width=\textwidth]{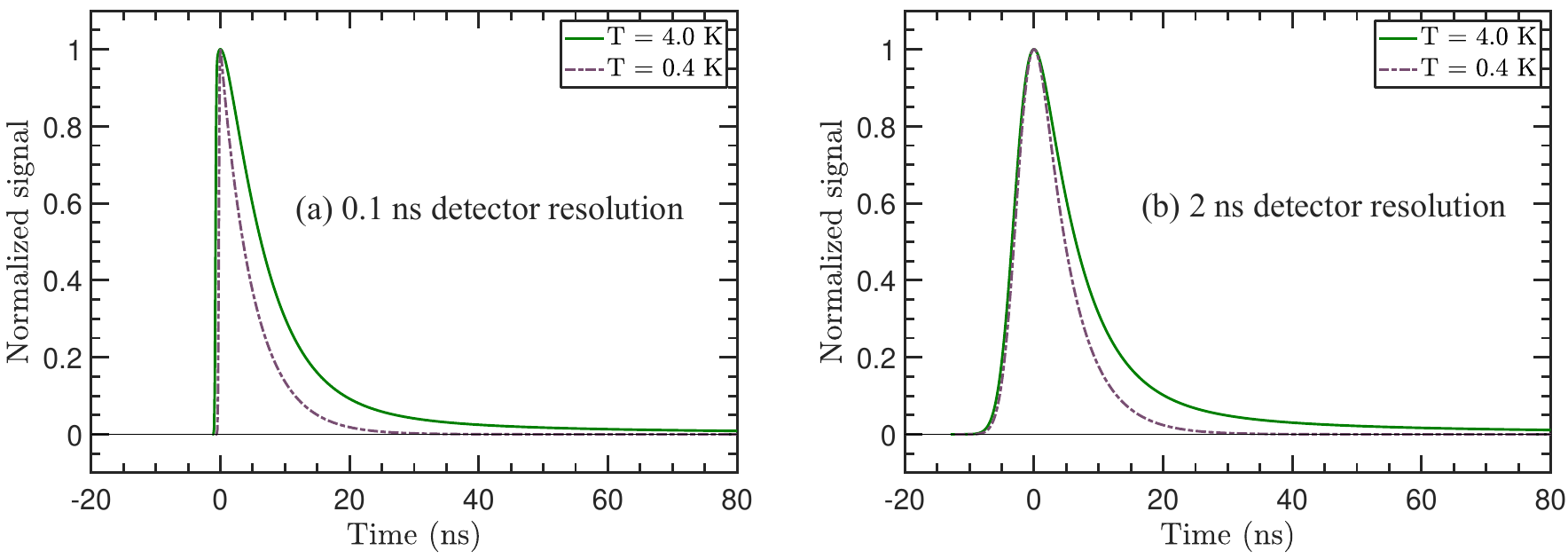}
		\caption{Calculations of the expected prompt scintillation signal pulse shape for electron ionization in LHe for temperature of 0.4~K and 4.0~K and for detector resolutions of (a) 0.1~ns and (b) 2~ns.  The singlet-state decay time constant is taken to be 5~ns.}
		\label{fig:simpulseshapes}
\end{figure*}

In LHe, the prompt scintillation signal is composed of two components, one from the recombination of ions leading to the production of molecular excimer states, and the other due to the neutral excitations that are created in the original ionization event.  Notably, before recombination occurs, the processes leading to the localization and thermalization of the ions to create snowball and bubble states happen very rapidly.  The timescale for these processes is of order a few 10s of ps \cite{Benderskii1999}, and once thermalized, these species drift toward each other to recombine.  As shown in this work, the recombination timescale is temperature dependent and can be longer than 100~ns.  For this reason, in the following discussion we will neglect the time required for the localization and thermalization of the ions.  It then follows that the critical timescale, besides the recombination time, affecting the shape of the prompt scintillation signal and which is independent of both the temperature and the type of ionizing particle is the singlet state mean lifetime, $\tau_{d}$.  This implies that regardless of how quickly recombination proceeds, the singlet state lifetime will represent an irreducible time resolution for any measurement.  

In practical applications, the experimental quantity of interest for particle detection and identification is the rate at which the signal is generated rather than the survival probability of charges, a quantity that has thus far been the focus of our discussion.  The signal shape is defined by the probability of emission of scintillation light within a time interval $t$ and $t + dt$ and temperature $T$.  We denote this probability as $S(t,T)dt$ with
\begin{equation} \label{eq:signalshape1}
S(t,T) =  \int_{0}^{\infty}  R(t, t-t')  F(t-t',T)dt'
\end{equation}
where
\begin{equation} \label{eq:signalshape2}
F(t, T) =  f_{r}\frac{\partial}{\partial t}\left[1-\overbar{\Omega}(t, T) \right] \ast H(t)  + f_{ex}H(t).  
\end{equation}
Here, $f_{r}$ and $f_{ex}$ are the fraction of the prompt scintillation signal due to recombination and excitation, respectively, and $H(t) = e^{-{t/\tau_{d}}}/\tau_{d} $ is the rate of scintillation emission from singlet state decays.  The component of the prompt signal due to recombination is the convolution of the rate of recombination, $\partial(1 -\overbar{\Omega}(t))/\partial t$ and the rate of decay of the singlets, $H(t)$, that are created by the recombination process.  The second term in Eq.~\ref{eq:signalshape2} is the contribution to the signal from the radiative decay of singlet state excimers produced from neutral excited atoms in the initial ionization process.  Both terms are convolved with the function, $R(t, t')$, which represents the detector response at time $t$ to an input at time $t'$, to produce the measured signal.  The contribution from instrument noise is neglected.

Measurements by Habicht~\cite{HabichtThesis} of the scintillation pulse shape for signals generated by ionizing radiation in LHe at 1.8~K show a decay time constant of $<7$~ns.  This, however, should represent an upper bound on the singlet-state decay lifetime because the effects of recombination are not negligible at this temperature.  Therefore, we assume a singlet state mean lifetime of 5~ns and use Eqs.~\ref{eq:signalshape1} and \ref{eq:signalshape2} to calculate the expected prompt scintillation signal from electron energy loss in helium at temperatures of 0.4~K and 4.0~K.  The detector response function is taken to be Gaussian with widths of 0.1 and 2~ns, where the former represents a near-ideal but still very much achievable experimental resolution with the use of fast response photo-detectors such as SiPMs.  The results of these calculations are shown in Fig.~\ref{fig:simpulseshapes}.  Absent from these figures are the calculated pulse shapes for $\alpha$-particle-induced scintillation.  However, if $\alpha$ recombination happens on a very fast timescale as suggested by Ref.\cite{Ito2012}, then the pulse shape will be comparable to that for electrons at 0.4~K and will be primarily determined by the singlet state lifetime.  If this presumption holds true, then a LHe detector operated around a temperature of 4~K will be able to observe a difference in the prompt pulse shape of the scintillation induced by an $\alpha$ particle versus that from a fast electron.

The measured prompt scintillation pulse shape generated by $\alpha$ particles and 1~MeV electrons in LHe at 4.2~K and 1.8~K made by Habicht~\cite{HabichtThesis} do show that there is, indeed, a difference between the pulses generated by these two types of particles at 4.2~K. This difference, however, appears to vanish at 1.8~K.  But altogether, their observations are consistent with our discussion of the temperature dependence of recombination.

There are several important aspects that must be considered in implementing this idea for particle identification.  First, the energy of the electron needs to be sufficiently high that the particle is truly minimally ionizing and the description of geminate recombination applies.  More precisely, the thermalization distribution must have a sufficiently heavy tail so that the effects of recombination are more easily detected.  Moreover, to accurately characterize the scintillation pulse shape, a sufficient number of photons needs to be detected, and processes that can further distort the time profile of the photons must be keep to a minimum.  One such process is the conversion of the EUV scintillation light to longer wavelengths by wavelength shifting fluors such as TPB.  The choice of wavelength shifter must then be guided by the need for both high conversion efficiency and fast decay time, but if possible, the removal of the wavelength shifter from the light detection process is most preferred. Another process that must also be suppressed is multiple scattering of photons as it would further degrade the timing resolution of the measurement.

Missing from the discussion thus far is the presence of an applied electric field which is commonly utilized in particle detectors employing liquid and gaseous scintillators as the detection medium.  This allows for a secondary signal (S2), in addition to the primary scintillation signal (S1), to be observed by extracting the ions onto a charge readout.  The purpose of this is to make use of the parameter $S2/S1$ for particle discrimination.  However, the collection of charges produced by a heavy ionizing particle in LHe is difficult.  For instance, at a field of 10~kV/cm, less than 10\% of the charges from an $\alpha$-particle is collected \cite{Ito2012}.  Whenever the application requires large-scale detectors, the voltages needed on electrodes are immensely high and may not be achievable in practice.  If the scintillation pulse shape (S1) can solely provide for the necessary discrimination power, then this potential problem may be circumvented, and detector design and operation are greatly simplified.  In case the electric field is necessary for the application, the pulse shape of S1 may be used in conjunction with $S1/S2$ to further enhance discrimination power.

A very different possibility is to use helium in the gaseous rather than liquid phase. In such a situation, the $S2$ signal can be measured through either charge avalanche amplification or proportional electroluminescence.  The latter is more advantageous from both an energy resolution and electrical stability standpoint.  Furthermore, the singlet state lifetime in helium gas at a pressure of 700~Torr is determined to be 0.55~ns~\cite{Hill1988}, which should help in distinguishing the effect of the radiative decay of singlet states relative to the recombination time on the pulse shape.  A more comprehensive consideration of this possibility, however, is beyond the scope of this work.

For temperatures below the $\lambda$ transition, exploiting the effects of recombination time for particle identification and discrimination  seems less workable given that recombination proceeds in similar swift fashion for both electrons and $\alpha$ particles.  However, this should not be entirely ruled out without definitive experimental evidence.  The zero-field prompt scintillation yield data in Sec.~\ref{sec:promptpes} hint at the possibility that the temperature-dependent recombination effect for $\alpha$ particles may in fact be the inverse of electrons, though this is merely speculation and requires further study for clarification.  Future measurements of the temperature and field dependence of the prompt scintillation pulse shape produced by the stopping of $\alpha$ particles and electrons in LHe will help answer this question.

Finally, considering that free electrons also form bubble or localized states in liquid neon~\cite{Kuper1961, Jortner1965, Springett1967, Bruschi1972, Loveland1972} as it does in LHe, the effects of recombination time may be also applicable to particle discrimination in liquid neon detectors.  Previous studies of particle discrimination in liquid neon~\cite{Nikkel2008, Lippincott2012} have been made but in the context of measuring the ratio of the prompt to delayed scintillation components.  The latter is analogous to the afterpulses observed in LHe scintillation.  Whether discrimination can be obtained through only the shape of the prompt pulse and what operational parameters (pressure, temperature, field, etc.) will maximize it are open questions worth considering.  Answers to these, however, are left to future work.

\section{\label{sec:conclusion}Conclusion}
The prompt scintillation signal generated by the passage of $\approx 364$~keV electrons in LHe at 0.44~K exhibits a $\approx 42$\% reduction at a field of 40~kV/cm.  We showed that the apparent temperature dependence of this reduction for electrons can be explained by an effect due to finite ion recombination time and signal integration time.  To the best of our knowledge, this is the first time that the effects of recombination time have been used to explain such an observation in noble liquids. The observation of this effect indicates the existence of a heavy-tailed distribution of thermalization distances produced by electrons as has been suggested by the work of Seidel~\textit{et al.}~\cite{Seidel2014}.  Furthermore, this thermalization distribution appears to have a dependence on the energy of the primary electron, with higher-energy electrons producing a heavier-tailed distribution.  A potential application of the recombination time effect is the use of pulse shape analysis for particle identification and discrimination in particle/nuclear physics experiments.

\begin{acknowledgments}
The authors greatly appreciate the help provided by the following
individuals and organizations: E.~Bond (LANL, C-NR) for
electroplating the electrode with radioactive sources, M.~Febbraro
(ORNL) for coating the lightguide with TPB, and LANSCE Facility
Operations for providing support for the experimental activities. This
work was supported by the U.S. Department of Energy, Office of Science,
Office of Nuclear Physics, through Contract No. 89233218CNA00000
(LANL) under proposal number LANLEEDM, Contract No. DE-AC05-
00OR22725 (ORNL), and Contract No. DE-SC0019309 (ASU).

\end{acknowledgments}


\end{document}